\documentclass[artical, twocolumn, twoside]{IEEEtran}

\IEEEoverridecommandlockouts

\usepackage[cmex10]{amsmath}
\usepackage{amssymb}
\usepackage{graphicx}
\usepackage{subfigure}
\usepackage{cite}
\usepackage{subeqnarray}
\usepackage{diagbox}
\usepackage{booktabs}
\usepackage{multirow}
\usepackage{multicol}
\usepackage{algorithm}
\usepackage{algorithmic}
\usepackage{setspace}
\usepackage{url}
\usepackage{xcolor}
\usepackage{array}
\usepackage{footnote}
\usepackage{enumerate}
\usepackage{framed}
\usepackage{lineno}
\usepackage{threeparttable}
\makesavenoteenv{tabular}

\newcommand{\mb}{\mathbf}

\newcommand{\mc}{\mathcal}
\newcommand{\ms}{\mathsf}
\newcommand{\mr}{\mathrm}

\newcommand{\tb}{\textbf}

\newtheorem{theorem}{Theorem}
\newtheorem{lemma}{Lemma}
\newtheorem{proposition}[theorem]{Proposition}

\newtheorem{definition}{Definition}

\newtheorem{remark}{Remark}

\begin{document}

\title{Dynamic Programming for Sequential Deterministic Quantization of Discrete Memoryless Channels}

\author{%
\IEEEauthorblockN{Xuan He, Kui Cai, Wentu Song, and Zhen Mei}\\
\thanks{Part of this work has been presented in ISIT 2019 \cite{	he2019dynamicISIT}.}
\thanks{This work is supported by RIE2020 Advanced Manufacturing and Engineering (AME) programmatic grant A18A6b0 057 and Singapore Ministry of Education Academic Research Fund Tier 2 MOE2019-T2-2-123.}
}

\maketitle

\begin{abstract}
In this paper, under a general cost function $C$, we present a dynamic programming (DP) method  to obtain an optimal sequential deterministic quantizer (SDQ) for $q$-ary input discrete memoryless channel (DMC).
The DP method has complexity $O(q (N-M)^2 M)$, where $N$ and $M$ are the alphabet sizes of the DMC output and quantizer output, respectively.
Then, starting from the quadrangle inequality, two techniques are applied to reduce the DP method's complexity.
One technique makes use of the Shor-Moran-Aggarwal-Wilber-Klawe (SMAWK) algorithm and achieves complexity $O(q (N-M) M)$.
The other technique is much easier to be implemented and achieves complexity $O(q (N^2 - M^2))$.
We further derive a sufficient condition under which the optimal SDQ is optimal among all quantizers and the two  techniques are applicable.
This generalizes the results in the literature for binary-input DMC.
Next, we show that the cost function of $\alpha$-mutual information ($\alpha$-MI)-maximizing quantizer belongs to the category of $C$.
We further prove that under a weaker condition than the sufficient condition we derived, the aforementioned two techniques are applicable to the design of $\alpha$-MI-maximizing quantizer.
Finally, we illustrate the particular application of our design method to practical pulse-amplitude modulation systems.
\end{abstract}

\begin{IEEEkeywords}

$\alpha$-mutual information, discrete memoryless channel, dynamic programming, quadrangle inequality, sequential deterministic quantizer.

\end{IEEEkeywords}

\IEEEpeerreviewmaketitle


\section{Introduction}

\begin{figure}[t]
\centering
\includegraphics[scale = 0.5]{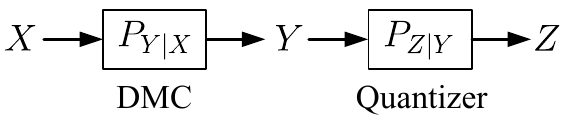}
\caption{Quantization of a discrete memoryless channel (DMC).}
\label{fig: DMC}
\end{figure}

Consider the quantization problem for the $q$-ary input discrete memoryless channel (DMC) with $q \geq 2$, as shown by Fig. \ref{fig: DMC}.
The channel input $X$ takes values from $\mc{X}$,
\begin{align*}
   \mc{X} = \{x_1, x_2, \ldots, x_q\},
\end{align*}
with probability
\[
    P_X(x_i) = \mr{Pr}(X = x_i) > 0, i \in [q],
\]
where $[n] = \{ 1, 2, \ldots, n\}$ for any positive integer $n$.
The channel output $Y$ takes values from $\mc{Y}$,
\begin{align*}
   \mc{Y} = \{y_1, y_2, \ldots, y_N\},
\end{align*}
with channel transition probability
\begin{align*}
   P_{Y|X}(y_j | x_i) = \mr{Pr}(Y &= y_j | X = x_i), i \in [q], j \in [N],
\end{align*}
where  $P_{Y|X}(y_j | x_i) \in [0, 1]$ and $\sum_{j \in [N]} P_{Y|X}(y_j | x_i) = 1$.
We assume $P_Y(y_j) = \sum_{i \in [q]} P_X(x_i) P_{Y|X}(y_j|x_i) > 0, \forall j \in [N]$ throughout the paper.
The most generic task is to design a quantizer
\[
    Q: \mc{Y} \to \mc{Z} = \{1, 2, \ldots, M\}
\]
to minimize a certain cost function $C(Q)$, where $2 \leq M < N$ is of interest.
Clearly, the quantizer $Q$ is uniquely specified by $P_{Z|Y}$, $Z$'s probability distribution conditioned on $Y$.

A deterministic quantizer (DQ) $Q: \mc{Y} \to \mc{Z}$ means that for each $y \in \mc{Y}$, there exists a unique $z' \in \mc{Z}$ such that 
\[
    P_{Z|Y}(z|y) =
    \begin{cases}
        1, & z = z',\\
        0, & z \neq z',
    \end{cases}
\]
or equivalently, we say $y$'s quantization result $Q(y)$ is a deterministic element in $\mc{Z}$.
For the cost function $C$ considered in this paper, we show that there always exists at least one DQ that is optimal among all quantizers.
Due to this reason as well as that DQ is more practical than non-deterministic quantizer, we focus only on DQs in this paper.
For any DQ $Q: \mc{Y} \to \mc{Z}$, denote $Q^{-1}(z) \subset \mc{Y}$ as the preimage of $z \in \mc{Z}$.

For binary-input DMC, dynamic programming (DP) \cite[Section 15.3]{introAlgo01} was applied by Kurkoski and Yagi \cite{Kurkoski14} to design quantizers that maximize the mutual information (MI)  between $X$ and $Z$, i.e., $I(X; Z)$.
The complexity (refer to the computational complexity throughout this paper unless the storage complexity is specified) of this DP method was reduced \cite{Iwata14, Sakai17} by applying the Shor-Moran-Aggarwal-Wilber-Klawe (SMAWK) algorithm \cite{Aggarwal87}.
However, for the general $q$-ary input DMC with $q > 2$, design of the optimal quantizers that maximize $I(X; Z)$ is an NP-hard problem  \cite{Nazer17, Laber18}.
Up till now, only the necessary condition \cite{Burshtein92, coppersmith1999partitioning}, rather than any sufficient condition, has been established for the optimal quantizer; meanwhile, there only exist some suboptimal design methods in practice \cite{Kurkoski08, Sakai14, Zhang16, Hassanpour17, Laber18}.

In this paper, we are not going to solve the NP-hard problem: finding an optimal DQ to minimize $C$ for a general $q$-ary input DMC.
Instead, we consider the optimal design of a specific type of DQ $Q: \mc{Y} \to \mc{Z}$ satisfying
\begin{equation}\label{eqn: definition of SDQ}
\left\{
\begin{array}{l}
    Q^{-1}(1) \,\,\,= \{{y_{\lambda_0 + 1}, y_2, \ldots, y_{\lambda_1}}\},\\
    Q^{-1}(2) \,\,\,= \{y_{\lambda_1 + 1}, y_{\lambda_1 + 2},  \ldots, y_{\lambda_2}\},\\
    \quad\quad\quad\quad\vdots\\
    Q^{-1}(M) = \{y_{\lambda_{M-1} + 1}, y_{\lambda_{M-1} + 2},  \ldots, y_{\lambda_M}\},
\end{array}
\right.
\end{equation}
where $0 = \lambda_0 < \lambda_1 < \lambda_2 < \cdots < \lambda_{M-1} < \lambda_M = N$.
We name this type of DQ sequential deterministic quantizer (SDQ).
The design of SDQs is called sequential deterministic quantization in this paper.
Based on \eqref{eqn: definition of SDQ}, every SDQ $Q: \mc{Y} \to \mc{Z}$ can be equivalently described by the integer set $\Lambda = \{\lambda_0 = 0, \lambda_1, \lambda_2, \ldots, \lambda_M = N\}$, in which each element is regarded as a quantization boundary/threshold.
Due to its simplicity, SDQ is generally more preferable in practical communication and data storage systems which usually have real-valued channel outputs.
For example, SDQs are used for  additive white Gaussian noise (AWGN) channels with quadrature amplitude modulations (QAMs) with Gray mappings \cite{lewandowsky2017message} (this channel model can essentially be  decomposed into AWGN channels with pulse-amplitude modulations (PAMs)), for non-volatile memory (NVM) channels which are similar to AWGN channels with PAMs \cite{aslam2016read, mei2019onchannel, mei2020deep, Wang14}, and also for hardware-friendly decoders of low-density parity-check (LDPC) codes  \cite{he2019mutual, he2019onfinite}.
These practical applications motivate us to investigate the design of SDQs for $q$-ary input DMCs, particularly the DMCs derived from AWGN channels with PAMs.

\subsection{Contributions of This Paper}

The main contributions of this work are summarized as follows.
We remark that since our results are generally for $q \geq 2$ and for a general cost function, they are non-trivial extensions of the results of \cite{Kurkoski14, Iwata14, Sakai17}.
\begin{enumerate}
\item   Under a general cost function $C$, we present a DP method with complexity $O(q (N-M)^2 M)$ to obtain an optimal SDQ.
\item   For the case where the quadrangle inequality (QI) \cite{	Yao80} holds,  we apply two techniques to reduce the complexity of the DP method.
    One technique achieves complexity $O(q (N-M) M)$ by making use of the SMAWK algorithm \cite{	Aggarwal87}, and the other technique is much easier to be implemented and achieves complexity $O(q (N^2 - M^2))$.
\item   We derive a sufficient condition under which the optimal SDQ is an optimal DQ and the above two low-complexity techniques are applicable.
\item   We make special effort design the $\alpha$-mutual information ($\alpha$-MI)\cite{Verdu15, Ho15, Csiszar95} maximizing SDQs. (In particular, for $\alpha = 1$, the $\alpha$-MI is exactly the standard MI, which is the most popular metric for channel quantization.)
    We show that the related cost function belongs to the category of $C$; consequently, the results mentioned in the first three contributions are also applicable here.
    Moreover, we prove that the two low-complexity techniques are actually applicable to the design of $\alpha$-MI-maximizing SDQs under a condition which is weaker than the sufficient condition mentioned in the third contribution.
\item   We investigate the quantization of DMCs derived from AWGN channels with PAMs.
    We illustrate that the weaker condition mentioned in the fourth contribution holds for this case.
    The numerical results demonstrate that the DP method optimized by the two low-complexity techniques can be much more efficient in terms of actual running time.
    Moreover, the optimal SDQs obtained by our DP method are better (have lower cost) than the DQs obtained by both the  greedy combining  algorithm \cite{Kurkoski08, Sakai14} and the Kullback-Leibler (KL)-means algorithm \cite{Zhang16}.
\end{enumerate}

\subsection{Organization}

The remainder of this paper is organized as follows.
Section \ref{section: preliminaries} presents some preliminaries for the quantizer design.
Section \ref{section: DP} develops a DP method for the sequential deterministic quantization of $q$-ary input DMCs.
Section \ref{section: reduce complexity} introduces the QI and applies two techniques to reduce the DP method's complexity.
Section \ref{section: MI quantizer} investigates the design of $\alpha$-MI-maximizing quantizer in details.
Section \ref{section: PAM} presents the numerical results for the quantization of DMCs derived from AWGN channels with PAMs.
Finally, Section \ref{section: conclusion} concludes the paper.

\subsection{Notations}

We list notations used throughout this paper.
$\mc{X} = \{x_1, x_2, \ldots, x_q\}$, $\mc{Y} = \{y_1, y_2, \ldots, y_N\}$, and $\mc{Z} = \{1, 2, \ldots, M\}$ denote the alphabets of the DMC input, DMC output, and quantizer output.
Their corresponding random variables are denoted by $X$, $Y$, and $Z$, whose realizations are denoted by $x$, $y$, and $z$, respectively.
The distributions or joint distributions of $X$, $Y$, and $Z$ are denoted in the conventional style, e.g., $P_X$, $P_{X, Y}$, $P_{Y|X}$, etc.

$Q$ denotes a DQ (possibly also an SDQ), while $\Lambda$ always denotes an SDQ.
$C$ denotes the cost function.
For $1 \leq i \leq j \leq N$, $w(i, j)$ denotes the cost caused by quantizing $\{y_i, y_{i+1}, \ldots, y_j\}$ into one level.
For $1 \leq m \leq n \leq N$, $\ms{dp}(n, m)$ denotes the minimum cost for using an SDQ to quantize $\{y_1, y_{2}, \ldots, y_n\}$ into $m$ levels.
For $m-1 \leq t < n$, denote $\ms{dp}_t(n, m) = \ms{dp}(t, m-1) + w(t+1, n)$ and $\ms{sol}(n, m) = \arg\min_{m-1 \leq t < n} \ms{dp}_t(n, m)$.

Let $\mathbb{R}$ (resp. $\mathbb{R}_+$) denote the (resp. nonnegative) real number set, and $[0, 1]$ denote the set of real numbers between 0 and 1 (both inclusive).
For any positive integer $n$, $[n] = \{1, 2, \ldots, n\}$.
Denote the $(q-1)$-dimensional probability simplex by
\[
\mc{U} = \{(a_1, a_2, \ldots, a_q) \mid a_1 + \cdots + a_q = 1, a_i \geq 0, i \in [q] \}.
\]
For any $\mb{a} = (a_i)_{1 \leq i \leq n}, \mb{b} = (b_i)_{1 \leq i \leq n} \in \mathbb{R}_+^n$, define the binary relation $\succeq$ between $\mb{a}$ and $\mb{b}$ by
\begin{equation}\label{eqn: >=}
    \mb{a} \succeq \mb{b} \iff a_i b_j \geq a_j b_i, \forall 1 \leq i < j \leq n.
\end{equation}

\section{Preliminaries}\label{section: preliminaries}

In this paper, for any quantizer $Q: \mc{Y} \to \mc{Z}$, we consider the following general cost function:
\begin{equation}\label{eqn: cost function}
    C(Q) = \sum_{z \in \mc{Z}} P_Z(z) \phi(P_{X|Z}(\cdot | z)),
\end{equation}
where $P_{X|Z}(\cdot | z) = (P_{X|Z}(x_1 | z), \ldots, P_{X|Z}(x_q | z))\in \mc{U}$ and $\phi: \mc{U} \to \mathbb{R}$ is concave on $\mc{U}$, i.e.,
\[
    \phi(t u_1 + (1-t) u_2) \geq t \phi(u_1) + (1-t) \phi(u_2)
\]
for any $u_1, u_2 \in \mc{U}$ and $t \in [0, 1]$.
Here, $C(Q)$ given by \eqref{eqn: cost function} is a general cost function used for minimum impurity partition in learning theory \cite{Burshtein92, coppersmith1999partitioning, Laber18}.
The minimum impurity partition problem is somewhat equivalent to the problem of finding the optimal quantizers for DMCs \cite{Kurkoski14}.
$C(Q)$ includes many popular concrete cost functions as subcases.
For example, $\phi(P_{X|Z}(\cdot | z)) = - \sum_{x \in \mc{X}} P_{X|Z}(x|z) \log P_{X|Z}(x|z)$ yields $I(X; Z) = H(X) - C(Q)$; as a result, $C(Q)$ becomes a valid cost function for MI-maximizing quantizer.
Later in Section \ref{section: MI quantizer}, we will also illustrate  that the cost function of $\alpha$-MI-maximizing quantizer belongs to the category of $C(Q)$.

\begin{lemma}\label{lemma: deterministic}
There exists a DQ $Q^*: \mc{Y} \to \mc{Z}$ which is optimal among all quantizers quantizing $\mc{Y}$ to $\mc{Z}$, i.e.,
\[
   C({Q}^*) = \min_{Q: \mc{Y} \to \mc{Z}} C(Q)
\]
with $C(Q)$ given by \eqref{eqn: cost function}.
\end{lemma}

\begin{IEEEproof}
See Appendix \ref{appendix: deterministic}.
\end{IEEEproof}

Lemma \ref{lemma: deterministic} generalizes \cite[Lemma 1]{Kurkoski14} since it uses  a general cost function.
It explains why we only consider DQ in this paper.
When only DQ is considered for \eqref{eqn: cost function}, we have
\begin{align*}
    &P_Z(z) \phi(P_{X|Z}(\cdot | z))
    \\=& \mr{Pr}(Y \in Q^{-1}(z)) \phi\left( \frac{\sum_{y \in Q^{-1}(z)} P_{X|Y}(\cdot|y) P_Y(y)}{\mr{Pr}(Y \in Q^{-1}(z))} \right),
\end{align*}
which can be considered as the (weighted) cost for quantizing $Q^{-1}(z)$ into one level.
Moreover, we have
\begin{align*}
    C(Q) &\geq \sum_{z \in \mc{Z}} \sum_{y \in Q^{-1}(z)} P_Y(y) \phi\left( P_{X | Y}(\cdot | y) \right)
    \\&= \sum_{y \in \mc{Y}} P_Y(y) \phi\left( P_{X | Y}(\cdot | y) \right),
\end{align*}
where the inequality is due to the concavity of $\phi$.
This implies that any quantizer cannot have a smaller cost than that before quantization, which indeed is  reasonable.

Denote
\begin{equation}\label{eqn: Delta}
    \Delta = \{\delta_1, \delta_2, \ldots, \delta_N\},
\end{equation}
where for $j \in [N]$, $\delta_j$ is given by
\[
    \delta_j = (P_{X|Y}(x_1|y_j), P_{X|Y}(x_2|y_j), \ldots, P_{X|Y}(x_q|y_j)),
\]
which can be regarded as a point in $\mc{U}$ from the viewpoint of geometry.
In this way, we establish a one-to-one mapping between $\Delta$ and $\mc{Y}$.
Define an equivalent quantizer of $Q: \mc{Y} \to \mc{Z}$ by
\[
    \tilde{Q}: \Delta \to \mc{Z}.
\]
They are equivalent in the sense that $\tilde{Q}(\delta_j) = Q(y_j)$ for $1 \leq j \leq N$ and $C(\tilde{Q}) = C(Q)$.
We have the following result.

\begin{lemma}\label{lemma: necessary condition for optimality}
There exists an optimal quantizer $\tilde{Q}^*: \Delta \to \mc{Z}$, i.e.,
\[
   C(\tilde{Q}^*) = \min_{\tilde{Q}: \Delta \to \mc{Z}} C(\tilde{Q}),
\]
such that $\tilde{Q}^*$ is deterministic and for any $z, z' \in \mc{Z}$ with $z \neq z'$, there exists a hyperplane that separates $\tilde{Q}^{*-1}(z)$ and $\tilde{Q}^{*-1}(z')$.
Moreover, the equivalent quantizer of $\tilde{Q}^*$, $Q^*: \mc{Y} \to \mc{Z}$, is also deterministic and optimal.
\end{lemma}

We omit the proof, since it is almost identical to the proof of \cite[Lemma 2]{Kurkoski14} except that a more general cost function is considered here.

For the binary-input case (i.e., $q = 2$), \cite{Kurkoski14} proved that if $P_{Y|X}$ satisfies
\begin{align*}
    &{P_{Y|X}(y_j|x_1)}{P_{Y|X}(y_{j+1}|x_2)} \geq \nonumber
    \\&\quad\quad\quad\quad {P_{Y|X}(y_{j+1}|x_1)}{P_{Y|X}(y_j|x_2)}, \forall j \in [N-1],
\end{align*}
any optimal SDQ is an optimal DQ that can maximize $I(X; Z)$.
Also, \cite{Kurkoski14} developed a DP method with complexity $O((N-M)^2 M)$ to obtain the optimal SDQ.
This DP method's complexity was reduced to $O((N-M)M)$ in \cite{Iwata14 } by applying the SMAWK algorithm  \cite{Aggarwal87}.
Moreover, \cite{Sakai17} further extended the result of \cite{Iwata14 } to $\alpha$-MI-maximizing quantizer.
These works motivate us to apply DP to obtain an optimal SDQ for a general $q$-ary input DMC.

\section{Dynamic Programming for Sequential Deterministic Quantization}
\label{section: DP}

In this section, we first present a DP algorithm for obtaining an optimal SDQ, and then derive a sufficient condition which ensures the global optimality of the optimal SDQ among all DQs.

\subsection{Dynamic Programming Algorithm}

For $1 \leq l \leq r \leq N$, denote $w(l, r)$ as the cost for quantizing $\{y_l, y_{l+1}, \ldots, y_{r}\}$ into one level, i.e.,
\begin{align}\label{eqn: w}
    w(l, r)
    = \sum_{j' = l}^{r} P_Y(y_{j'}) \phi\left( \frac{\sum_{j = l}^{r} P_{X,Y}(\cdot, y_j)}{\sum_{j'' = l}^{r} P_Y(y_{j''})} \right).
\end{align}
To simplify the computation of $w(\cdot, \cdot)$, we precompute and store $\sum_{j = 1}^{k} {P_{X, Y}}(x_i, y_j)$ for $k = 1, 2, \ldots, N$ and $i = 1, 2, \ldots, q$, both the computational and storage complexities of which are $O(q N)$ (Hence this term does not dominate the complexities of the algorithms  discussed later in the paper).
In this case, we can generally compute $w(l, r)$ with a computational complexity linear to the input alphabet size $q$.
We thus denote the computational complexity for computing $w(l, r)$ for any given pair of $(l, r)$ by $O(q)$.
Note that we can also precompute and store $w(\cdot, \cdot)$, with computational and storage complexities of $O(q N^2)$ and $O(N^2)$, respectively.
However, this is not necessary since we can compute $w(l, r)$ on-the-fly for any pair of $(l, r)$ with computational complexity $O(q)$ when needed.
The computational complexity of each algorithm discussed later in the paper is given for the case where $w(\cdot, \cdot)$ is not precomputed.
When the precomputation is applied, an algorithm's computational complexity may change and is lower-bounded by $O(q N^2)$, with an extra storage complexity of $O(N^2)$.
As an example, we will discuss this situation for the DP method proposed later in this section.

For $1 \leq m \leq n \leq N$, let $\Lambda(n, m) = \{\lambda_0, \lambda_1, \ldots, \lambda_m\}, 0 = \lambda_0 < \lambda_1 < \cdots < \lambda_m = n$ be an SDQ for quantizing $\{y_1, y_{2}, \ldots, y_{n}\}$ into $m$ levels.
We have
\begin{equation*}
    C(\Lambda(n, m)) = \sum_{i = 1}^{m} w(\lambda_{i - 1} + 1, \lambda_i).
\end{equation*}
Moreover, let
\[
    \Lambda^*(n, m) = \{\lambda^*_0, \lambda^*_1, \ldots, \lambda^*_m\} = \arg\min_{\Lambda(n, m)} C(\Lambda(n, m)).
\]
Our task is to obtain a $\Lambda^*(N, M)$.
Recall that
\begin{align*}
    \ms{dp}(n, m) &= C(\Lambda^*(n, m)),
\\    \ms{dp}_t(n, m) &= \ms{dp}(t, m-1) + w(t+1, n), m-1 \leq t < n,
    \\\ms{sol}(n, m) &= \arg\min_{m-1 \leq t < n} \ms{dp}_t(n, m).
\end{align*}
Algorithm \ref{algo: DP N2M} summarizes the computation of $\Lambda^*(N, M)$.

\begin{proposition}\label{theorem: DP with N2M}
Algorithm \ref{algo: DP N2M} is correct and runs in $O(q (N-M)^2 M)$ time.
\end{proposition}

\begin{IEEEproof}
For $m = 1$, we have $\ms{dp}(n, m) = w(1, n)$.
For $m > 1$, we have
\begin{align*}
\ms{dp}(n, m) &= \sum_{i = 1}^{m} w(\lambda^*_{i - 1} + 1, \lambda^*_i)
\\&= \ms{dp}(\lambda^*_{m-1}, m-1) +  w(\lambda^*_{m-1} + 1, n)
\\&= \ms{dp}_{\ms{sol}(n, m)}(n, m),
\end{align*}
implying that Algorithm \ref{algo: DP N2M} is correct.

On the other hand, clearly, the computational complexity of Algorithm \ref{algo: DP N2M} is dominated by the computation between lines \ref{code: DP N2M @ for m} and \ref{code: DP N2M @ end for m}, which is $O(q (N-M)^2 M)$.
(Recall that this complexity is given for the case where $w(\cdot, \cdot)$ is not precomputed. It becomes $O(q N^2 + (N-M)^2 M)$ when $w(\cdot, \cdot)$ is precomputed.)
\end{IEEEproof}

\begin{algorithm}[t!]
\caption{Dynamic programming for obtaining $\Lambda^*(N, M)$}
\label{algo: DP N2M}
\begin{algorithmic}[1]
\REQUIRE $P_X, P_{Y|X}, N, M$.
\ENSURE $\Lambda^*(N, M)$.

\STATE  $//$\textit{Initialization}
\FOR    {$n \leftarrow 1, 2, \ldots, N$}
    \STATE  $\ms{dp}(n, 1) \leftarrow w(1, n)$.
    \STATE  $\ms{sol}(n, 1) \leftarrow 0$.
\ENDFOR

\STATE  $//$\textit{Compute} $\ms{dp}(N, M)$
\FOR    {$m \leftarrow 2, 3, \ldots, M$}\label{code: DP N2M @ for m}
    \FOR    {$n \leftarrow N-M+m, N-M-1+m, \ldots, m$}\label{code: DP N2M @ for n}
        \STATE  $\ms{sol}(n, m) \leftarrow \arg\min_{m-1 \leq t < n} \ms{dp}_t(n, m)$.\label{code: algo 1 @ {sol} min_t}
        \STATE  $\ms{dp}(n, m) \leftarrow \ms{dp}_{\ms{sol}(n, m)}(n, m)$.
    \ENDFOR\label{code: DP N2M @ end for n}
\ENDFOR\label{code: DP N2M @ end for m}

\STATE  $//$\textit{Recursively generate $\Lambda^*(N, M)$}
\STATE  $\lambda^*_{M} \leftarrow N$.
\FOR    {$m \leftarrow M, M-1, \ldots, 1$}
    \STATE  $\lambda^*_{m-1} \leftarrow \ms{sol}(\lambda^*_{m}, m)$.
\ENDFOR

\RETURN $\Lambda^*(N, M)$.
\end{algorithmic}
\end{algorithm}


\subsection{A Sufficient Condition}

We now derive a sufficient condition under which the optimal SDQ obtained by Algorithm \ref{algo: DP N2M} is an optimal DQ (and thus is also optimal among all quantizers).
We assume that there exist at least two points $\delta_j, \delta_{j'} \in \Delta$  defined by \eqref{eqn: Delta} such that $\delta_j \neq \delta_{j'}$; otherwise, any DQ will have the same cost value according to \eqref{eqn: cost function}.
Consider the situation where all points in $\Delta = \{\delta_1, \delta_2, \ldots, \delta_N\}$ are located on a line, i.e., there exists a unique $t_j \in \mathbb{R}$ for any $\delta_j \in \Delta$ such that
\begin{equation}\label{eqn: line}
    \delta_j = \delta_1 + t_j \mb{d}, ~~\mb{d} = (d_i)_{1 \leq i \leq q} = \delta_N - \delta_1,
\end{equation}
where the addition and substraction are element-wise and we assume $\delta_1 \neq \delta_{N}$ (since we can replace $\delta_N$ by any $\delta_j \neq \delta_1$).
$\delta_1, \delta_2, \ldots, \delta_N$ are said to sequentially located on a line if and only if we further have
\begin{equation}\label{eqn: line sequentially}
    0 = t_1 \leq t_2 \leq \cdots \leq t_N = 1.
\end{equation}
We have the following result.

\begin{theorem}\label{theorem: line to optimal}
The following three statements are equivalent:
\begin{enumerate}
\item   $\delta_1, \delta_2, \ldots, \delta_N$ (defined by \eqref{eqn: Delta}) are sequentially located on a line.
\item   $\delta_1, \delta_2, \ldots, \delta_N$ are located on a line, and the elements in $\mc{X}$ can be relabelled to make $P_{Y|X}$ satisfy
\begin{align}\label{eqn: (i, j) > (i', j')}
    P_{Y|X}(\cdot|x_{i}) \succeq P_{Y|X}(\cdot|x_{i'}), \forall 1 \leq i < i' \leq q,
\end{align}
where $\succeq$ is defined in \eqref{eqn: >=}.
\item   $\delta_1, \delta_2, \ldots, \delta_N$ are located on a line, and the elements in $\mc{X}$ can be relabelled to make $P_{Y|X}$ satisfy
\begin{align}\label{eqn: (i, j) > (i+1, j+1)}
    &{P_{Y|X}(y_j|x_{i})}{P_{Y|X}(y_{j+1}|x_{i+1})} \geq \nonumber
    \\& \quad\quad\quad\quad\quad\quad{P_{Y|X}(y_{j+1}|x_{i})}{P_{Y|X}(y_j|x_{i+1})}, \nonumber
    \\& \quad\quad\quad\quad\quad\quad\quad\quad\, \forall i \in [q-1], j \in [N-1].
\end{align}
\end{enumerate}
Moreover, if $\delta_1, \delta_2, \ldots, \delta_N$ are sequentially located on a line, any optimal SDQ is an optimal DQ.
\end{theorem}

\begin{IEEEproof}
See Appendix \ref{appendix: line to optimal}.
\end{IEEEproof}

Note that if $\delta_1, \delta_2, \ldots, \delta_N$ are located on a line, we can always make them sequentially located on the line by relabelling the elements in $\mc{Y}$.
Specifically, denote $t_{j_1}, t_{j_2}, \ldots, t_{j_N}$ as the result after sorting $t_1, t_2, \ldots, t_N$ given by \eqref{eqn: line} in ascending order.
After relabelling $y_{j_k}$ as $y_k$ for $k \in [N]$, $\delta_1, \delta_2, \ldots, \delta_N$ (corresponding to the new labelling) are sequentially located on the line.
We also show in \eqref{eqn: di+1 > di}  how to further relabel the elements in $\mc{X}$ to  make $P_{Y|X}$ satisfy \eqref{eqn: (i, j) > (i', j')} and \eqref{eqn: (i, j) > (i+1, j+1)}.
Moreover, for the binary-input case (i.e., $q = 2$), $\delta_1, \delta_2, \ldots, \delta_N$ are always located on a line.
In such a case, the elements in $\mc{Y}$ can always be relabelled to make $P_{Y|X}$ satisfy \eqref{eqn: (i, j) > (i', j')} and \eqref{eqn: (i, j) > (i+1, j+1)}, after which any optimal SDQ is an optimal DQ.
This situation is fully investigated in \cite{Kurkoski14} for the MI-maximizing quantizer, while being included as a subcase of our results.

\section{Reducing the Complexity of Dynamic Programming}\label{section: reduce complexity}

In certain cases the DMC output alphabet size $N$ can be very large, and hence Algorithm \ref{algo: DP N2M}  may need to take a long time to find an optimal solution.
For example, when we use Algorithm \ref{algo: DP N2M} to quantize the output of a continuous memoryless channel to $M$ levels, we may need to first uniformly quantize the continuous output to $N$ levels, after which Algorithm \ref{algo: DP N2M} can be applied.
Obviously, increasing $N$ can reduce the loss due to uniform quantization.
Thus, it is worth reducing the computational complexity of Algorithm \ref{algo: DP N2M} to make it work well for large $N$.

We develop two low-complexity techniques in this section.
Both techniques rely on the QI which is defined as follows.
The QI was first proposed by Yao \cite{Yao80} as a sufficient condition to reduce the complexity of a class of DP.
Then, it was pointed out in \cite{Bein09} that Yao's result  can be achieved by using the SMAWK algorithm \cite{Aggarwal87}.

\begin{definition}[Quadrangle inequality]
$w(\cdot, \cdot)$ (see \eqref{eqn: w}) is said to satisfy the QI if it satisfies
\begin{equation}\label{eqn: QI}
    w(i, k) + w(j, l) \leq w(i, l) + w(j, k)
\end{equation}
for all $1 \leq i < j \leq k < l \leq N$.
\end{definition}

\subsection{First Technique: SMAWK Algorithm}

Inspired by the works of \cite{	Bein09} and \cite{Iwata14},  for $2 \leq m \leq M$, we define $D^{m} = [d^{m}_{i, j}]_{1 \leq i, j \leq N-M+1}$ as a matrix with $d^{m}_{i, j}$ given by
\begin{equation}\label{eqn: dij}
    d^{m}_{i, j} =
    \begin{cases}
        \ms{dp}_{j-2+m}(i-1+m, m), & i \geq j,\\
        \infty, & i < j,
    \end{cases}
\end{equation}
where $\infty$ indeed can be replaced by any constant larger than all $d^{m}_{i, j}$ for $i \geq j$.

We define $D^{m}$ as above since it can be computed in the order of $D^2, D^3, \ldots, D^M$, and for $m \leq n \leq N-M+m$, $\ms{dp}(n, m)$ is given by the minima of the $(n-m+1)$-th row of $D^{m}$.
More specifically, for a given $m$, let $\mb{p} = (p_i)_{1 \leq i \leq N-M+1}$, where $p_i$ is the position (column index) of the leftmost minima in the $i$-th row of $D^m$, i.e., $p_i$ is the smallest integer such that $d^m_{i, p_i} = \min_{j \in [N-M+1]} d^m_{i, j}$.
Then, we have
\begin{align*}
    \ms{sol}(n, m) &= \arg\min_{m-1 \leq t \leq n-1} \ms{dp}_t(n, m)
    \\&= p_{n-m+1} - 2 + m.
\end{align*}
As a result, the computation in lines \ref{code: DP N2M @ for n} to \ref{code: DP N2M @ end for n} of Algorithm \ref{algo: DP N2M} corresponds to the new problem of computing $\mb{p}$.
The new problem is essentially the classical problem discussed in \cite{Aggarwal87}, where the SMAWK algorithm was proposed to solve this problem when $D^{m}$ is totally monotone.

\begin{definition}[Totally monotone matrix]
A $2 \times 2$ matrix $A = [a_{i, j}]_{1 \leq i, j \leq 2}$ is monotone if $a_{1, 1} > a_{1, 2}$ implies $a_{2, 1} > a_{2, 2}$.
A matrix $D$ is totally monotone if every $2 \times 2$ submatrix (intersections of arbitrary two rows and two columns) of $D$ is monotone.
\end{definition}

Assume $D^m$ is totally monotone.
The SMAWK algorithm for finding the leftmost minima in each row of $D^m$ is summarized in Algorithm \ref{algo: SMAWK}.
For a subset of rows $\mb{r}$ and columns $\mb{c}$ of $D^m$, let $D^m(\mb{r}, \mb{c})$ denote the submatrix of $D^m$ which consists of the intersections of rows $\mb{r}$ and columns $\mb{c}$.
The function $\mr{SMAWK}(\mb{r}, \mb{c})$ is to find the column indices of the leftmost minima in each row of  $D^m(\mb{r}, \mb{c})$, and the function $\mr{Reduce}(\mb{r}, \mb{c})$ is to reduce $D^m(\mb{r}, \mb{c})$  to size $|\mb{r}| \times |\mb{r}|$ by deleting $|\mb{c}| - |\mb{r}|$ many ``dead" columns in which the leftmost minima are not located.
The essential ideas of both $\mr{SMAWK}(\mb{r}, \mb{c})$ and $\mr{Reduce}(\mb{r}, \mb{c})$  are to make use of the total monotonicity of $D^m(\mb{r}, \mb{c})$.
We further remark that Algorithm \ref{algo: SMAWK} does not require $D^{m}$ to be precomputed, but a specific entry of  $D^{m}$ to be computed on-the-fly when needed.
According to \eqref{eqn: dij}, any entry $d^{m}_{i, j}$ of $D^{m}$ can be computed with the same complexity as computing $w(j-1+m, i-1+m)$, i.e., $O(q)$ in general.
Therefore, the total complexity of Algorithm \ref{algo: SMAWK} is $O(q (N-M))$ \cite{Aggarwal87}.
More details about Algorithm \ref{algo: SMAWK} can be found in \cite{Aggarwal87}.

\begin{algorithm}[!t]
\caption{SMAWK algorithm for finding the leftmost minima in each row of $D^{m}$}
\label{algo: SMAWK}
\begin{algorithmic}[1]
\STATE  $\mb{r} \leftarrow \mb{c} \leftarrow (1, 2, \ldots, N-M+1)$.
\RETURN $\mr{SMAWK}(\mb{r}, \mb{c})$.

\STATE
\STATE  \tb{Function:} $\mr{SMAWK}(\mb{r}, \mb{c})$
\STATE  $\mb{c} \leftarrow \mr{Reduce}(\mb{r}, \mb{c})$.
\IF {$|\mb{r}| = 1$}
    \RETURN $\mb{p} \leftarrow \mb{c}$.
\ELSE
    \STATE  $\mb{r}' \leftarrow (r_2, r_4, \ldots, r_{\lfloor |\mb{r}|/2 \rfloor \cdot 2})$.
    \STATE  $(p_2, p_4, \ldots, p_{\lfloor |\mb{r}|/2 \rfloor \cdot 2}) \leftarrow \mr{SMAWK}(\mb{r}', \mb{c})$.
    \STATE  $j \leftarrow 1$.
    \FOR    {$i = 1, 3, \ldots, \lceil |\mb{r}|/2 \rceil \cdot 2 - 1$}
        \STATE  $p_i \leftarrow c_j$.
        \STATE  If $i < |\mb{r}|$, $u \leftarrow p_{i+1}$; otherwise, $u \leftarrow \infty$.
        \WHILE  {$j \leq |\mb{r}|$ and $c_j \leq u$}
            \STATE  $p_i \leftarrow c_j$ if $d^m_{r_i, c_j} < d^m_{r_i, p_i}$.
            \STATE  $j \leftarrow j + 1$.
        \ENDWHILE
        \STATE  $j \leftarrow j - 1$.
    \ENDFOR
    \RETURN $\mb{p} \leftarrow (p_i)_{1 \leq i \leq |\mb{r}|}$.
\ENDIF

\STATE
\STATE  \tb{Function:} $\mr{Reduce}(\mb{r}, \mb{c})$
\STATE  $i \leftarrow 1$.
\WHILE {$|\mb{r}| < |\mb{c}|$}
    \IF {$d^m_{r_i, c_i} \leq d^m_{r_i, c_{i+1}}$ and $i < |\mb{r}|$}
        \STATE  $i \leftarrow i + 1$.
    \ELSIF {$d^m_{r_i, c_i} \leq d^m_{r_i, c_{i+1}}$ and $i = |\mb{r}|$}
        \STATE  Delete $c_{i+1}$ from $\mb{c}$.
    \ELSIF {$d^m_{r_i, c_i} > d^m_{r_i, c_{i+1}}$}
        \STATE  Delete $c_{i}$ from $\mb{c}$.
        \STATE  $i \leftarrow i - 1$ if $i > 1$.
    \ENDIF
\ENDWHILE
\RETURN $\mb{c}$. 

\end{algorithmic}
\end{algorithm}


The following lemma illustrates the connection between the QI and the total monotonicity.
It implies that if $w(\cdot, \cdot)$ satisfies the QI, the complexity of Algorithm \ref{algo: DP N2M} can be reduced to $O(q (N-M) M)$ by applying Algorithm \ref{algo: SMAWK}.

\begin{lemma}\label{lemma: QI monotone}
If $w(\cdot, \cdot)$ satisfies the QI, $D^{m}$ is totally monotone for $2 \leq m \leq M$.
\end{lemma}

\begin{IEEEproof}
Consider the $2 \times 2$ submatrix of $D^{m}$  consisting of the intersections of rows $k, l$ with $k < l$ and columns $i, j$ with $i < j$, denoted by $D_s$.
If $j > k$, we have $d^{m}_{k, j} = \infty$, implying $D_s$ is monotone.
For $j \leq k$, we have $d^{m}_{k, i} + d^{m}_{l, j} - d^{m}_{k, j} - d^{m}_{l, i} = w(i+m, k+m) + w(j+m, l+m) - w(j+m, k+m) - w(i+m, l+m) < 0$ because $w(\cdot, \cdot)$ satisfies the QI.
Then, $d^{m}_{k, i} > d^{m}_{k, j}$ implies $d^{m}_{l, i} > d^{m}_{l, j}$, indicating $D_s$ is monotone.
This completes the proof.
\end{IEEEproof}

\subsection{Second Technique}

To check whether $w(\cdot, \cdot)$ satisfies the QI is vital for reducing the complexity of Algorithm \ref{algo: DP N2M}.
For $w(\cdot, \cdot)$ that cannot be determined analytically of whether it satisfies the QI or not, we can test it by  exhaustively checking \cite{Bein09}
\begin{equation}\label{eqn: check QI at N^2}
    w(r, s) + w(r+1, s+1) \leq w(r, s+1) + w(r+1, s)
\end{equation}
for $1 \leq r < s < N$.
It can be easily proved that \eqref{eqn: QI} is equivalent to \eqref{eqn: check QI at N^2}.
Checking \eqref{eqn: check QI at N^2} has complexity $O(q N^2)$, which will lower-bound the overall complexity for the quantizer design if it is applied.
It is worth doing the checking if $q N^2 < q (N-M)^2 M$, i.e., the checking costs less complexity than Algorithm \ref{algo: DP N2M}.

If $w(\cdot, \cdot)$ is verified by the exhaustive test to satisfy the QI, the SMAWK algorithm can be used to reduce the complexity of Algorithm \ref{algo: DP N2M}, and hence the overall complexity approaches the lower-bound of $O(q N^2)$.
Considering that implementing the SMAWK algorithm is tricky and sophisticated, in the following, we present another low-complexity DP algorithm which is much easier to be implemented, and the overall complexity also approaches this lower-bound.
By simply modifying the upper and lower bounds of $t$ in line \ref{code: algo 1 @ {sol} min_t} of Algorithm \ref{algo: DP N2M} (i.e. the standard DP algorithm), it can reduce the complexity from $O(q(N-M)^2M)$ to $O(q(N^2-M^2))$.
The corresponding details are as follows.

\begin{lemma}\label{lemma: {sol}}
If $w(\cdot, \cdot)$ satisfies the QI, we then have
\begin{equation}\label{eqn: {sol}}
    \ms{sol}(n, m-1) \leq \ms{sol}(n, m) \leq \ms{sol}(n+1, m)
\end{equation}
for $2 \leq m \leq n < N$.
\end{lemma}

\begin{IEEEproof}
See Appendix \ref{appendix: proof of {sol}}.
\end{IEEEproof}

The inequality of \eqref{eqn: {sol}} was first proved by Yao as a consequence of the QI in order to reducing the complexity for solving the DP problem considered in \cite{Yao80}.
Though our DP problem is different from that considered  in \cite{Yao80}, fortunately, \eqref{eqn: {sol}} still holds as a consequence of the QI and can also be used to reduce the complexity for solving our DP problem.
In particular, when \eqref{eqn: {sol}} holds, for $n = N-M+m-1, N-M+m-2, \ldots, m$ in line \ref{code: DP N2M @ for n} of Algorithm \ref{algo: DP N2M}, we can conduct a low-complexity technique by enumerating $t$ in line \ref{code: algo 1 @ {sol} min_t} from $\max\{m-1, \ms{sol}(n, m-1)\}$ to $\min\{n-1, \ms{sol}(n+1, m)\}$ instead of from $m-1$ to $n-1$.
Let $T(n, m)$ denote the complexity for enumerating $t$ in line \ref{code: algo 1 @ {sol} min_t} with respect to the $m$ in line \ref{code: DP N2M @ for m} and the $n$ in line \ref{code: DP N2M @ for n}.
Then, the total complexity for enumerating $t$, after applying this low complexity algorithm, is given by
\begin{align*}
    &\sum_{m = 2}^{M} \sum_{n = m}^{N-M+m} T(n, m)\\
    \leq&  \sum_{m = 2}^{M} T(N-M+m, m) + \\
    &\quad \sum_{m = 2}^{M} \sum_{n = m}^{N-M+m-1} (\ms{sol}(n+1, m) - \ms{sol}(n, m-1) + 1)\\
    \leq& M (N-M+1) + \sum_{n = M+1}^{N} \ms{sol}(n, M)\\
    \leq& (N+M)(N-M+1)
\end{align*}
Therefore, this low-complexity algorithm has complexity $O(q (N^2 - M^2))$.

\subsection{Remarks}

The two low-complexity techniques presented in this section can be used to reduce the complexity of Algorithm \ref{algo: DP N2M} once $w(\cdot, \cdot)$ satisfies the QI, no matter this requirement is verified analytically or by exhaustive test.
The first technique making use of the SMAWK algorithm works faster, while being more complicated than the second technique making use of \eqref{eqn: {sol}} in terms of the implementation complexity.

\begin{theorem}\label{theorem: line to reduce complexity}
If $\delta_1, \delta_2, \ldots, \delta_N$ defined by \eqref{eqn: Delta} are sequentially located on a line, $w(\cdot, \cdot)$ satisfies the QI.
\end{theorem}

\begin{IEEEproof}
See Appendix \ref{appendix: line to reduce complexity}.
\end{IEEEproof}

Theorem \ref{theorem: line to optimal} together with Theorem \ref{theorem: line to reduce complexity} indicate that, if $\delta_1, \delta_2, \ldots, \delta_N$ are located on a line, we can first relabel the elements in $\mc{Y}$ to make $\delta_1, \delta_2, \ldots, \delta_N$ sequentially located on the line.
Then, an optimal DQ can be obtained by using the DP method given by Algorithm \ref{algo: DP N2M}, and at the same time, the two low-complexity techniques become applicable.
This result extends the results of \cite{Kurkoski14,	Iwata14, Sakai17 } to cases with $q > 2$ and to a more general cost function.

\section{$\alpha$-Mutual Information-Maximizing Quantizer}
\label{section: MI quantizer}

In this section, we consider a specific quantizer, the $\alpha$-MI-maximizing quantizer for $\alpha > 0$.
The $\alpha$-MI  is closely related to Gallager's exponent function \cite{gallager1965simple} and to the channel capacity problems in many  applications (e.g., see \cite{polyanskiy2010arimoto}).
In particular, it can be used to measure the channel capacity of order $\alpha$ \cite{Csiszar95}.
For $\alpha > 0$, the $\alpha$-MI between $X$ and $Z$ is defined by \eqref{eqn: alpha-MI} \cite{Verdu15, Ho15, Sakai17, Csiszar95}.
\begin{figure*}[!t]
    \begin{equation}\label{eqn: alpha-MI}
        I_{\alpha}(X; Z) =
        \begin{cases}
            \sum_{z \in \mc{Z}} \sum_{x \in \mc{X}} P_{X, Z}(x, z)  \log \frac{P_{X, Z}(x, z)}{P_X(x)P_Z(z)}, & \alpha = 1,\\
            \log \left( \sum_{z \in \mc{Z}} \max_{x \in \mc{X}} P_{Z|X}(z|x) \right), & \alpha = \infty,\\
            \frac{\alpha}{\alpha - 1} \log \left( \sum_{z \in \mc{Z}} \left( \sum_{x \in \mc{X}} P_X(x) P_{Z|X}^{\alpha}(z|x) \right)^{1/\alpha} \right), & \alpha \in (0, 1) \cup (1, \infty).
        \end{cases}
    \end{equation}
\end{figure*}
Note that $I_1(X; Z)$ is equivalent to the standard MI between $X$ and $Z$, i.e., $I(X; Z)$, and $I_{1/2}(X; Z)$ is equivalent to the cutoff rate between $X$ and $Z$ \cite{Csiszar95}.

We first illustrate that the cost function of an $\alpha$-MI-maximizing quantizer can be defined as a specific case of \eqref{eqn: cost function}.
To this end, for $\alpha > 0$, define the cost function of any quantizer $Q: \mc{Y} \to \mc{Z}$ by
\begin{equation}\label{eqn: cost function alpha}
    C_{\alpha}(Q) = \sum_{z \in \mc{Z}} P_Z(z) \phi_{\alpha}(P_{X|Z}(\cdot | z)),
\end{equation}
where
\begin{align*}
    &\phi_{\alpha}(P_{X|Z}(\cdot | z)) = \\&\quad\quad
    \begin{cases}
        - \sum_{x \in \mc{X}} P_{X|Z}(x|z) \log P_{X|Z}(x|z), & \alpha = 1,\\
        - \max_{x \in \mc{X}} P_{X|Z}(x|z) / P_X(x), & \alpha = \infty,\\
        \left( \sum_{x \in \mc{X}} P_X^{1 - \alpha}(x) P_{X|Z}^{\alpha}(x|z) \right)^{1/\alpha}, &  \alpha \in (0, 1),\\
        - \left( \sum_{x \in \mc{X}} P_X^{1 - \alpha}(x) P_{X|Z}^{\alpha}(x|z) \right)^{1/\alpha}, &  \alpha \in (1, \infty).
    \end{cases}
\end{align*}
The cost function $C_{\alpha}(Q)$ given by \eqref{eqn: cost function alpha} is a specific case of that given by \eqref{eqn: cost function}, since it can be easily proved that $\phi_{\alpha}: \mc{U} \to \mathbb{R}$ is concave on $\mc{U}$ (e.g., see \cite[Lemma 1]{Sakai17}).
On the other hand, we have
\begin{equation}\label{eqn: I-alpha = C-alpha}
    I_{\alpha}(X; Z) =
    \begin{cases}
        H(X) - C_{\alpha}(Q), & \alpha = 1,\\
        \log(- C_{\alpha}(Q)), & \alpha = \infty,\\
        \frac{\alpha}{\alpha - 1} \log \left( C_{\alpha}(Q)  \right), &  \alpha \in (0, 1),\\
        \frac{\alpha}{\alpha - 1} \log (- C_{\alpha}(Q)), &  \alpha \in (1, \infty),
    \end{cases}
\end{equation}
where $H(X) = -\sum_{x \in \mc{X}} P_X(x) \log P_X(x)$ is a constant given $P_X$.
According to \eqref{eqn: I-alpha = C-alpha}, maximizing $I_{\alpha}(X; Z)$ is equivalent to minimizing $C_{\alpha}(Q)$.
This implies that design of the $\alpha$-MI-maximizing quantizers belongs to the quantizer design category discussed in the previous sections, and hence all the previous results are applicable here.

We now consider the design of $\alpha$-MI-maximizing SDQs.
Since the cost function varies for different $\alpha$, to avoid ambiguity, $w(\cdot, \cdot)$ is now replaced by $w_{\alpha}(\cdot, \cdot)$, which can be computed based on \eqref{eqn: w} with $\phi(\cdot)$ being replaced by $\phi_{\alpha}(\cdot)$.
We have the following result.

\begin{theorem}\label{theorem: QI vs LLR increasing}
If the elements in $\mc{X}$ can be relabelled to make $P_{Y|X}$ satisfy \eqref{eqn: (i, j) > (i', j')}, $w_{\alpha}(\cdot, \cdot)$ satisfies the QI.
\end{theorem}

\begin{IEEEproof}
See Appendix \ref{appendix: proof of QI vs LLR increasing}.
\end{IEEEproof}

We remark that Theorem \ref{theorem: QI vs LLR increasing} does not require $\delta_1, \delta_2, \ldots, \delta_N$ to be located on a line, while Theorem \ref{theorem: line to reduce complexity} does.
In fact, the condition required by Theorem \ref{theorem: QI vs LLR increasing} is necessary, but not sufficient, for the condition required by Theorem \ref{theorem: line to reduce complexity} to hold, i.e., the condition that the elements in $\mc{X}$ can be relabelled to make $P_{Y|X}$ satisfy \eqref{eqn: (i, j) > (i', j')}  is necessary, but not sufficient, for $\delta_1, \delta_2, \ldots, \delta_N$ to be sequentially located on a line.
Therefore, when considering the design of $\alpha$-MI-maximizing SDQs, Theorem \ref{theorem: QI vs LLR increasing} is a stronger statement than Theorem \ref{theorem: line to reduce complexity} as it requires a weaker condition.
We show in the next section that Theorem \ref{theorem: QI vs LLR increasing} is applicable to the DMCs derived from AWGN channels with PAMs.
However, for other scenarios with $q > 2$, it is generally hard to relabel the elements in $\mc{X}$ (and even also in $\mc{Y}$) to make $P_{Y|X}$ satisfy \eqref{eqn: (i, j) > (i', j')}.

\section{Quantization of AWGN Channels with PAMs}\label{section: PAM}

\begin{figure}[t]
\centering
\includegraphics[scale = 0.5]{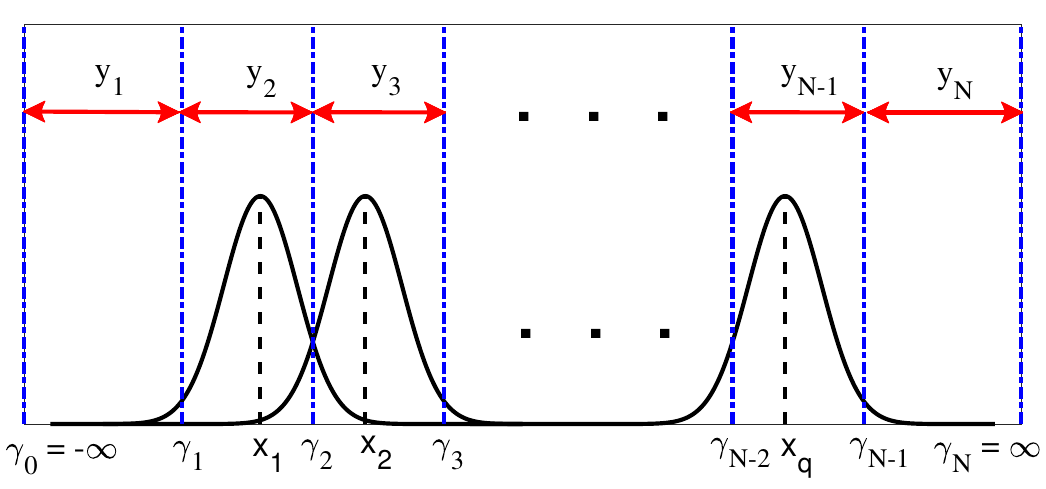}
\caption{PAM system with input $X \in \mc{X} = \{x_1, x_2, \ldots, x_q\}, x_1 < x_2 < \cdots < x_q$. $X$ is transmitted over an AWGN channel with noise variance $\sigma^2$ and mean 0. The channel is converted into a DMC by uniformly quantizing the continuous channel output $\widetilde{Y}$ to $N$ levels $\mc{Y} = \{y_1, y_2, \ldots, y_N\}$ based on $N+1$ thresholds $\Gamma = \{\gamma_0, \gamma_1, \ldots, \gamma_N\}$ with $\gamma_1 = x_1 - 3\sigma$, $\gamma_{N-1} = x_q + 3\sigma$, and $\gamma_1 - \gamma_2 = \gamma_2 - \gamma_3 = \cdots = \gamma_{N-2} - \gamma_{N-1}$.}
\label{fig: PAM}
\end{figure}

In this section, we consider the quantization of the PAM system shown  in Fig. \ref{fig: PAM}. The probability density function (pdf) of the channel continuous output $\widetilde{Y} = \widetilde{y}$ conditioned on channel input $X = x_i$ is given by
\[
    f_{\widetilde{Y}|X}(\widetilde{y}|x_i) = \frac{1}{\sqrt{2 \pi \sigma^2}} \exp \left({-\frac{(\widetilde{y} - x_i)^2}{2\sigma^2}} \right)
\]
for $i \in [q]$ and $\widetilde{y} \in \mathbb{R}$.
Our goal is to use an SDQ to quantize $\widetilde{Y}$ into $Z \in \mc{Z} = \{1, 2, \ldots, M\}$. That is, the quantization should be done by finding $M + 1$ thresholds $\Theta = \{\theta_0, \theta_1, \ldots, \theta_M\}, \theta_0 = -\infty < \theta_1 < \cdots < \theta_{M-1} < \theta_M = +\infty$, such that for $i \in [M]$, $\widetilde{Y} \in (\theta_{i-1}, \theta_{i}]$ is quantized to $Z = i$.

We first convert the channel into a DMC, as shown in Fig. \ref{fig: PAM}, with output $Y \in \mc{Y} = \{y_1, y_2, \ldots, y_N\}$, where $N \gg M$.
More specifically, we create $N + 1$ candidate thresholds $\Gamma = \{\gamma_0, \gamma_1, \ldots, \gamma_N\}, \gamma_0 = -\infty < \gamma_1 < \cdots < \gamma_{N-1} < \gamma_N = +\infty$, such that the transition probability of the DMC is given by
\[
    P_{Y|X}(y_j|x_i) =  \int_{\gamma_{j-1}}^{\gamma_{j}} f_{\widetilde{Y}|X}(\widetilde{y}|x_i) \mr{d} \widetilde{y}
\]
for $i \in [q]$ and $j \in [N]$.
In general, we can set $\gamma_1 = x_1 - 3\sigma$ and $\gamma_{N-1} = x_q + 3\sigma$, and set $\gamma_j = \gamma_{j-1} + (\gamma_{N-1} - \gamma_1) / (N-2)$ for $j = 2, \ldots, N-2$ to uniformly partition $[\gamma_1, \gamma_{N-1}]$ into $N - 2$ segments.

We can then use Algorithm \ref{algo: DP N2M} to find the optimal thresholds $\Theta$ from the candidate thresholds  $\Gamma$ according to Proposition \ref{theorem: DP with N2M}.
In particular, if $\alpha$-MI-maximizing SDQs are of interest, the two low-complexity techniques discussed in Section \ref{section: reduce complexity} are applicable here, according to Theorem \ref{theorem: QI vs LLR increasing} and the following lemma.
We remark that Lemma \ref{lemma: PAM QI} only requires $x_1 < x_2 < \cdots < x_q$ and $\gamma_0 = -\infty < \gamma_1 < \cdots < \gamma_{N-1} < \gamma_N = +\infty$.
However, it does not depend on $P_X$ and the specific values of $\{x_1, \ldots, x_q, \gamma_1, \ldots, \gamma_{N-1}\}$.
Moreover, Lemma \ref{lemma: PAM QI} also implies that if only $x_1 < x_2 < \cdots < x_q$ does not hold, $\{x_1, \ldots, x_q\}$ can be relabelled to make $P_{Y|X}$ satisfy \eqref{eqn: (i, j) > (i', j')} such that Theorem \ref{theorem: QI vs LLR increasing} is also applicable.

\begin{lemma}\label{lemma: PAM QI}
For the converted DMC shown in Fig. \ref{fig: PAM},  $P_{Y|X}$ satisfies \eqref{eqn: (i, j) > (i', j')}.
\end{lemma}

\begin{IEEEproof}
For $1 \leq i < q$ and $1 \leq j \leq N$, we have
\begin{align}\label{eqn: P_(Y|X)(yj/xi)}
    &P_{Y|X}(y_j|x_i) / P_{Y|X}(y_{j}|x_{i+1})\nonumber\\
    =& \frac{\int_{\gamma_{j-1}}^{\gamma_{j}} f_{\widetilde{Y}|X}(\widetilde{y}|x_i) \mr{d} \widetilde{y} }{ \int_{\gamma_{j-1}}^{\gamma_{j}} f_{\widetilde{Y}|X}(\widetilde{y}|x_{i+1}) \mr{d} \widetilde{y}}\nonumber\\
    =& \frac{ \int_{\gamma_{j-1}}^{\gamma_{j}} {f_{\widetilde{Y}|X}(\widetilde{y}|x_i)}\big/{f_{\widetilde{Y}|X}(\widetilde{y}|x_{i+1})} f_{\widetilde{Y}|X}(\widetilde{y}|x_{i+1}) \mr{d} \widetilde{y} }{ \int_{\gamma_{j-1}}^{\gamma_{j}} f_{\widetilde{Y}|X}(\widetilde{y}|x_{i+1}) \mr{d} \widetilde{y}}.
\end{align}
Since
\[
    \frac{f_{\widetilde{Y}|X}(\widetilde{y}|x_i)}{f_{\widetilde{Y}|X}(\widetilde{y}|x_{i+1})} = \exp \left(\frac{(x_i - x_{i+1}) (2\widetilde{y} - x_i - x_{i+1})}{ 2\sigma^2} \right)
\]
keeps strictly decreasing when $\widetilde{y}$ increases from $\gamma_{j-1}$ to $\gamma_j$, we have
\begin{equation}\label{eqn: f_{Y|X} decrease}
    \frac{f_{\widetilde{Y}|X}(\gamma_{j-1}|x_i)}{f_{\widetilde{Y}|X}(\gamma_{j-1}|x_{i+1})}
    >\frac{f_{\widetilde{Y}|X}(\widetilde{y}|x_i)}{f_{\widetilde{Y}|X}(\widetilde{y}|x_{i+1})}
    >\frac{f_{\widetilde{Y}|X}(\gamma_{j}|x_i)}{f_{\widetilde{Y}|X}(\gamma_{j}|x_{i+1})}
\end{equation}
for $\gamma_{j-1} < \widetilde{y} < \gamma_{j}$.
Then, based on \eqref{eqn: P_(Y|X)(yj/xi)} and \eqref{eqn: f_{Y|X} decrease}, we have
\[
    \frac{f_{\widetilde{Y}|X}(\gamma_{j-1}|x_i)}{f_{\widetilde{Y}|X}(\gamma_{j-1}|x_{i+1})}
    >\frac{P_{Y|X}(y_j|x_i)}{P_{Y|X}(y_j|x_{i+1})}
    >\frac{f_{\widetilde{Y}|X}(\gamma_{j}|x_i)}{f_{\widetilde{Y}|X}(\gamma_{j}|x_{i+1})},
\]
indicating that $P_{Y|X}$ satisfies \eqref{eqn: (i, j) > (i', j')}.
\end{IEEEproof}

\begin{figure}[t]
\centering
\includegraphics[scale = 0.5]{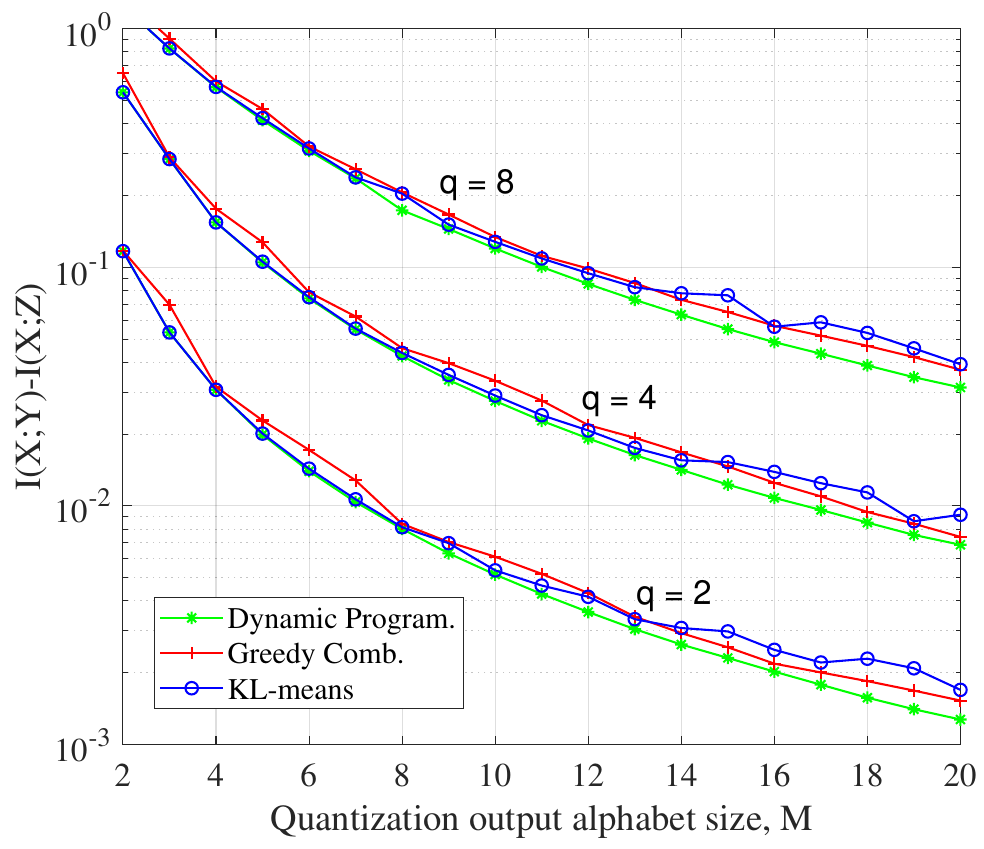}
\caption{Performance of the DP method, the greedy combining \cite{Kurkoski08, Sakai14}, and the KL-means \cite{Zhang16} algorithms.}
\label{fig: 248_PAM}
\end{figure}

Next, for the converted DMC shown in Fig. \ref{fig: PAM}, we compare the quantization performance  of the DP method with the prior art quantizer design algorithms proposed for the general $q$-ary input DMC, i.e., the greedy combining algorithm \cite{Kurkoski08, Sakai14} and the KL-means algorithm \cite{Zhang16}.
The MI-maximizing quantizers are of interest, and the MI gap $I_{g} = I(X; Y) - I(X; Z)$ is used as the comparison metric, which is the smaller the better.
In the simulations, we use uniform distribution for $P_X$.
We set $\sigma = 1$, $x_i = 2i - q - 1$ for $i \in [q]$, and $N = 128$.
$\mc{Y}$ is quantized to $M$ levels, $M = 2, 3, \ldots, 20$.
When the KL-means algorithm is used, we randomly choose $M$ out of $N$ points as the initial means (see \cite{Zhang16}) for $T_i = 100$ times, and for each time the KL-means algorithm  runs for $T_r = 100$ iterations to obtain a DQ, and finally the best ($I_{g}$ is minimized) DQ among the $T_i$ times is chosen.
The simulation results are illustrated by Fig. \ref{fig: 248_PAM}.
It is shown that the DP method performs better than both the greedy combining and KL-means algorithms, for different values of $q$.

Moreover, note that the two low-complexity techniques are applicable here.
The DP method has complexity $O(q (N-M) M)$ if applying the SMAWK algorithm and $O(q (N^2 - M^2))$ if applying \eqref{eqn: {sol}}.
In contrast, the greedy combining and KL-means algorithms have complexities $O(q N^2 (N-M))$ and $O(T_i T_r q N M)$, respectively, and hence are much more complex than the DP method.
As an example, we show the actual running time for these algorithms in Table \ref{table: running time}.

\begin{table}[!t]
\renewcommand{\arraystretch}{1.3}
\caption{Average Running Time on a Standard Desktop for Different Quantizer Design Algorithms, Where A1--A5 Refer to the Original DP Algorithm Given by Algorithm \ref{algo: DP N2M}, the DP Algorithm Optimized by Using the SMAWK Algorithm, the DP Algorithm Optimized by Using \eqref{eqn: {sol}}, the Greedy Combining Algorithm, and the KL-Means Algorithm, Respectively, and $q = 2, M = 8, T_i = 100, T_r = 100$ Are Used}
\label{table: running time}
\centering
    \begin{tabular}{clcc}
        \toprule
        \multirow{2}{1.5cm}{~~Algorithm} & \multirow{2}{1.5cm}{~~~~~~Complexity} & \multicolumn{2}{c}{Running time in second}\\\cline{3-4}
          & & $N = 128$ & $N = 1000$ \\

        \midrule
        A1 & $O(q (N-M)^2 M)$   & 0.042 & 2.323\\
        A2 & $O(q (N-M) M)$     & 0.004 & 0.045\\
        A3 & $O(q (N^2-M^2))$   & 0.007 & 0.349\\
        A4 & $O(q N^2 (N-M))$   & 0.206 & 89.496\\
        A5 & $O(T_i T_r q N M)$ & 1.353 & 10.063\\

        \bottomrule
    \end{tabular}
\end{table}

Our final remark is that the optimal SDQ for the converted DMC shown by Fig. \ref{fig: PAM} may not be a globally optimal DQ.
One toy counter-example has the following parameters: $(q, N, M, \sigma^2) = (4, 7, 4, 0.1)$, $(x_1, x_2, x_3, x_4) = (-3, -1, 1, 3)$, and $P_X = (0.53, 0.23, 0.23, 0.01)$.
We also find a counter-example  among millions of test cases with uniformly distributed $P_X$ and randomly generated $P_{Y|X}$ which satisfies \eqref{eqn: (i, j) > (i', j')}.
Both counter-examples imply that the condition of \eqref{eqn: (i, j) > (i', j')}  cannot solely guarantee the global optimality of an optimal SDQ among all DQs.
To guarantee the global optimality, one sufficient condition given by Theorem \ref{theorem: line to optimal} is to additionally require $\delta_1, \delta_2, \ldots, \delta_N$ to be located on a line.
An intriguing but very hard problem is to find a more general sufficient condition.

\section{Conclusion}
\label{section: conclusion}

In this paper, under the general cost function $C$ given by \eqref{eqn: cost function}, we have presented a DP method with  complexity $O(q (N-M)^2 M)$ to obtain an optimal SDQ for $q$-ary input DMC.
Two efficient techniques have been applied to reduce the DP method's complexity once $w(\cdot, \cdot)$ satisfies the QI.
One technique makes use of the SMAWK algorithm and achieves complexity $O(q (N-M) M)$.
The other one is much easier to be implemented and achieves complexity $O(q (N^2 - M^2))$.
We have proved that when $\delta_1, \delta_2, \ldots, \delta_N$ defined by \eqref{eqn: Delta} are sequentially located on a line, the optimal SDQ is an optimal DQ and the two efficient techniques are applicable.
This result generalizes the results of \cite{Kurkoski14, Iwata14, Sakai17}.
Next, we have showed that the cost function of an $\alpha$-MI-maximizing quantizer can be defined as a specific case of $C$.
We have further proved that if the elements in $\mc{X}$ can be relabelled to make $P_{Y|X}$ satisfy \eqref{eqn: (i, j) > (i', j')}, but not requiring $\delta_1, \delta_2, \ldots, \delta_N$ to be sequentially located on a line, the aforementioned two efficient techniques are applicable to the design of $\alpha$-MI-maximizing quantizer.
Finally, we have demonstrated the application of our design method to the DMCs derived from AWGN channels with PAMs.

\appendices

\section{Proof of Lemma \ref{lemma: deterministic}}\label{appendix: deterministic}

Note that for any quantizer $Q: \mc{Y} \to \mc{Z}$, $Q$ is specified by $P_{Z|Y}$, and $C(Q)$ given by \eqref{eqn: cost function} is a function of $P_{Z|Y}$.
We now show $C(Q)$ is concave on $P_{Z|Y}$.
For any $t \in [0, 1]$ and any two quantizers $Q^{(1)}, Q^{(2)}$ specified by $P^{(1)}_{Z|Y}, P^{(2)}_{Z|Y}$, respectively, denote $Q$ as the quantizer specified by $P_{Z|Y} = t P^{(1)}_{Z|Y} + (1-t) P^{(2)}_{Z|Y}$, where the addition is element-wise.
Then, for $x \in \mc{X}$ and $z \in \mc{Z}$, we have
\begin{align*}
    P_Z(z) &= t P^{(1)}_Z(z) + (1-t) P^{(2)}_Z(z),
    \\P_{X|Z}(x|z) &= \frac{t P^{(1)}_Z(z)}{P_Z(z)} P^{(1)}_{X|Z}(x|z) +
    \\&\quad\quad\quad\quad\quad\quad\frac{(1-t) P^{(2)}_Z(z)}{P_Z(z)} P^{(2)}_{X|Z}(x|z).
\end{align*}
Since $\phi$ is concave on $P_{X|Z}$, we have
\begin{align*}
    C(Q) &= \sum_{z \in \mc{Z}} P_Z(z) \phi(P_{X|Z}(\cdot | z))
    \\&\geq \sum_{z \in \mc{Z}} \big(t P^{(1)}_Z(z) \phi(P^{(1)}_{X|Z}(\cdot | z)) +
    \\&\quad\quad\quad\quad\quad\quad (1-t) P^{(2)}_Z(z) \phi(P^{(2)}_{X|Z}(\cdot | z))\big)
    \\&= t C(Q^{(1)}) + (1 - t) C(Q^{(2)}),
\end{align*}
indicating that $C(Q)$ is concave on $P_{Z|Y}$.
It is well known that there exists at least one extreme point, which corresponds to a DQ in this case, to make the concave function $C(Q)$ achieve its minima.
This completes the proof.

\section{Proof of Theorem \ref{theorem: line to optimal}}\label{appendix: line to optimal}

\emph{1) $\to$ 2):}
If $\delta_1, \delta_2, \ldots, \delta_N$ are sequentially located on a line, they are definitely located on the line and both \eqref{eqn: line} and \eqref{eqn: line sequentially} hold.
We relabel the elements in $\mc{X}$ to satisfy
\begin{equation}\label{eqn: di+1 > di}
    P_{X|Y}(x_{i}|y_1) d_{i+1} \geq P_{X|Y}(x_{i+1}|y_1) d_i, \forall i \in [q - 1],
\end{equation}
which is always possible.
For any $1 \leq i < i' \leq q$, if $P_{X|Y}(x_{i'}|y_1) = 0$, we have $d_{i'} > 0$ and
\begin{equation}\label{eqn: di' > di}
    P_{X|Y}(x_{i}|y_1) d_{i'} \geq P_{X|Y}(x_{i'}|y_1) d_i.
\end{equation}
If $P_{X|Y}(x_{i'}|y_1) > 0$, we have $P_{X|Y}(x_{i' - 1}|y_1) > 0$ due to \eqref{eqn: di+1 > di}.
Recursively, we have $P_{X|Y}(x_{k}|y_1) > 0$ for $i \leq k \leq i'$.
In this case, according to \eqref{eqn: di+1 > di}, we have
\[
    \frac{d_i}{P_{X|Y}(x_{i}|y_1)} \leq \frac{d_{i+1}}{P_{X|Y}(x_{i+1}|y_1)} \leq \cdots \leq \frac{d_{i'}}{P_{X|Y}(x_{i'}|y_1)},
\]
indicating that \eqref{eqn: di' > di} also holds.
Then, for $1 \leq i < i' \leq q$ and $1 \leq j < j' \leq N$, we have
\begin{align*}\label{eqn: P(yj|x) > P(yj+1|X)}
    &{P_{Y|X}(y_j|x_{i})}{P_{Y|X}(y_{j'}|x_{i'})} - {P_{Y|X}(y_{j'}|x_{i})}{P_{Y|X}(y_j|x_{i'})}\nonumber
    \\&=  {(P_{X|Y}(x_{i}|y_1) d_{i'} - P_{X|Y}(x_{i'}|y_1) d_i) (t_{j'} - t_j)} \cdot \nonumber
    \\&\quad\quad\quad\quad\quad\quad\quad\quad P_Y(y_j)P_Y(y_{j'})/P_X(x_i)/P_X(x_{i'}) \nonumber
    \\ &\geq 0,
\end{align*}
indicating that the second statement of Theorem \ref{theorem: line to optimal} is true.

\emph{2) $\to$ 3):}
This is a trivial conclusion.

\emph{3) $\to$ 1):}
Suppose that $\delta_1, \delta_2, \ldots, \delta_N$ are located on a line.
As a result, \eqref{eqn: line} holds.
Further assume that the elements in $\mc{X}$ can be relabelled to make $P_{Y|X}$ satisfy \eqref{eqn: (i, j) > (i+1, j+1)}, and we implement the relabelling in this way.
After that, for any $i \in [q-1]$ and $j \in [N-1]$, we have
\begin{align*}
    0 \leq &{P_{X|Y}(x_{i}|y_j)}{P_{X|Y}(x_{i+1}|y_{j+1})} -
    \\& \quad\quad\quad\quad\quad\quad{P_{X|Y}(x_{i}|y_{j+1})}{P_{X|Y}(x_{i+1}|y_j)}
    \\=& (P_{X|Y}(x_{i}|y_1) d_{i+1} - P_{X|Y}(x_{i+1}|y_1) d_i) (t_{j+1} - t_j).
\end{align*}
Then, for any $i \in [q-1]$, we have
\begin{align*}
    0 \leq &\sum_{j \in [N-1]} \big({P_{X|Y}(x_{i}|y_j)}{P_{X|Y}(x_{i+1}|y_{j+1})} -
    \\& \quad\quad\quad\quad\quad\quad{P_{X|Y}(x_{i}|y_{j+1})}{P_{X|Y}(x_{i+1}|y_j)} \big)
    \\=& P_{X|Y}(x_{i}|y_1) d_{i+1} - P_{X|Y}(x_{i+1}|y_1) d_i.
\end{align*}
As a result, if $t_{j+1} < t_j$ for some $j \in [N-1]$, we must have
\begin{equation}\label{eqn: di+1 = di}
    P_{X|Y}(x_{i}|y_1) d_{i+1} = P_{X|Y}(x_{i+1}|y_1) d_i, \forall i \in [q - 1].
\end{equation}

We now prove \eqref{eqn: di+1 = di} is not true.
If \eqref{eqn: di+1 = di} holds, we have $P_{X|Y}(x_{i}|y_1) > 0, \forall i \in [q]$;
otherwise, $P_{X|Y}(x_{i}|y_1) = 0, \forall i \in [q]$ can be derived but this is not true.
Then, we have
\[
    \frac{d_1}{P_{X|Y}(x_{1}|y_1)} = \frac{d_2}{P_{X|Y}(x_{2}|y_1)} = \cdots = \frac{d_q}{P_{X|Y}(x_{q}|y_1)}.
\]
Further since $0 = \sum_{i \in [q]} P_{X|Y}(x_{i}|y_N) - 1 = \sum_{i \in [q]} (P_{X|Y}(x_{i}|y_1) + t_N d_i) - 1 = \sum_{i \in [q]} d_i$, we must have $d_1 = d_2 = \cdots = d_q = 0$.
In this case, we have $0 = \mb{d} = \delta_N - \delta_1$, contradicting to the assumption that $\delta_1 \neq \delta_N$.
Therefore, \eqref{eqn: di+1 = di} is not true and hence we have $t_{j+1} \geq t_j, \forall j \in [N-1]$, indicating that $\delta_1, \delta_2, \ldots, \delta_N$ are sequentially located on the line.

\emph{Optimality:}
Suppose that $\delta_1, \delta_2, \ldots, \delta_N$ are sequentially located on a line.
According to Lemma \ref{lemma: necessary condition for optimality}, there exists an optimal DQ $\tilde{Q}^*: \Delta \to \mc{Z}$ such that for any $z, z' \in \mc{Z}$ with $z \neq z'$, there exists a point that separates $\tilde{Q}^{*-1}(z)$ and $\tilde{Q}^{*-1}(z')$ on the line.
In this case, the equivalent quantizer of $\tilde{Q}^*$, $Q^*: \mc{Y} \to \mc{Z}$, is an optimal DQ as well as an optimal SDQ.

\section{Proof of Lemma \ref{lemma: {sol}}}\label{appendix: proof of {sol}}

For $2 \leq m \leq n < N$, let $t = \ms{sol}(n, m)$ for brevity.
For any $m-1 \leq k < t$, we have
\begin{align*}
    &\ms{dp}_{t}(n+1, m) - \ms{dp}_k(n+1, m)\\
    =& \ms{dp}_t(n, m) - w(t+1, n) + w(t+1, n+1) - \\
    &\quad\quad (\ms{dp}_k(n, m) - w(k+1, n) + w(k+1, n+1))\\
    \leq& w(t+1, n+1) + w(k+1, n) - \\
    &\quad\quad w(t+1, n) - w(k+1, n+1)\\
    \leq& 0,
\end{align*}
where the last inequality holds because $w(\cdot, \cdot)$ satisfies the QI.
Therefore, we have $\ms{sol}(n, m) = t \leq \ms{sol}(n+1, m)$.

We now continue to prove $\ms{sol}(n, m) \geq \ms{sol}(n, m-1)$.
For $m = 2$, we have $\ms{sol}(n, m) \geq \ms{sol}(n, m-1) = 0$ trivially.
For $m \geq 3$, let $t = \ms{sol}(n, m-1)$ for brevity.
For any $m-1 \leq k < t$, we have
\begin{align*}
    &\ms{dp}_{t}(n, m) - \ms{dp}_k(n, m)\\
    =& \ms{dp}(t, m-1) + \ms{dp}_t(n, m-1) - \ms{dp}(t, m-2) - \\
    &\quad\quad (\ms{dp}(k, m-1) + \ms{dp}_k(n, m-1) - \ms{dp}(k, m-2))\\
    \leq& \ms{dp}(k, m-2) + \ms{dp}(t, m-1) - \\
    &\quad\quad \ms{dp}(k, m-1) - \ms{dp}(t, m-2).
\end{align*}
We continue the proof by first proving the following lemma.

\begin{lemma}\label{lemma: Monge {dp}}
For $2 \leq m \leq i < j \leq N$, denoting $\ms{dp}(i, m-1) + \ms{dp}(j, m) - \ms{dp}(i, m) - \ms{dp}(j, m-1)$ by $\psi(i, j, m)$, we have
\begin{equation*}
    \psi(i, j, m) \leq 0.
\end{equation*}
\end{lemma}

\begin{IEEEproof}
Let $t = \ms{sol}(i, 2)$ for brevity.
We have
\begin{align*}
    \psi(i, j, 2)
    &\leq \ms{dp}(i, 1) + \ms{dp}_t(j, 2) - \ms{dp}_t(i, 2) - \ms{dp}(j, 1)\\
    &= w(1, i) + w(t+1, j) - w(t+1, i) - w(1, j)\\
    &\leq 0.
\end{align*}
We then inductively prove $\psi(i, j, m) \leq 0$ for $m \geq 3$ given $\psi(i, j, m-1) \leq 0$ for $m-1 \leq i < j$.

Let $a = \ms{sol}(i, m)$ and $b = \ms{sol}(j, m-1)$ for brevity.
Note that $m - 1 \leq a < i$.
If $a < b$, we have
\begin{align*}
    &\psi(i, j, m)\\
    \leq& \ms{dp}_a(i, m-1) + \ms{dp}_b(j, m) - \ms{dp}_a(i, m) - \ms{dp}_b(j, m-1)\\
    =& \ms{dp}(a, m-2) + \ms{dp}(b, m-1) - \\
    &\quad\quad\quad\quad\quad\quad\quad\quad\quad\quad \ms{dp}(a, m-1) - \ms{dp}(b, m-2)\\
    =& \psi(a, b, m-1) \\
    \leq& 0.
\end{align*}
If $a \geq b$, we have
\begin{align*}
    &\psi(i, j, m)\\
    \leq& \ms{dp}_b(i, m-1) + \ms{dp}_a(j, m) - \ms{dp}_a(i, m) - \ms{dp}_b(j, m-1)\\
    =& w(b+1, i) + w(a+1, j) - w(a+1, i) - w(b+1, j)\\
    \leq& 0.
\end{align*}
This completes the proof of Lemma \ref{lemma: Monge {dp}}.
\end{IEEEproof}

At this point, we have $\ms{dp}_{t}(n, m) - \ms{dp}_k(n, m) \leq \psi(k, t, m-1) \leq 0$, implying $\ms{sol}(n, m) \geq \ms{sol}(n, m-1)$.

\section{Proof of Theorem \ref{theorem: line to reduce complexity}}\label{appendix: line to reduce complexity}

\begin{lemma}\label{lemma: concave on a line}
Let $n$ be a positive integer. $A, B, C, D \in \mathbb{R}^n$ are located on a line with $A$ and $D$ being the endpoints.
$\eta$ is a function which is concave on this line.
If there exist $\gamma, \beta \in [0, 1]$ such that $\gamma A + (1-\gamma)D = \beta B + (1-\beta) C$, we then have
\begin{align*}
    \gamma \eta(A) + (1-\gamma)\eta(D) \leq \beta \eta(B) + (1-\beta) \eta(C).
\end{align*}
\end{lemma}

\begin{IEEEproof}
If $A = D$, Lemma \ref{lemma: concave on a line} holds.
For $A \neq D$, there exist unique $\theta, \tau \in [0, 1]$ such that
\begin{align*}
    B &= \theta A + (1-\theta)D,
    \\ C &= \tau A + (1-\tau) D.
\end{align*}
Then, we have
\begin{align*}
    &\gamma A + (1-\gamma)D
    \\=& \beta B + (1-\beta) C
    \\=& (\beta \theta + (1-\beta) \tau) A + (\beta (1-\theta) + (1-\beta) (1-\tau)) D,
\end{align*}
which leads to
\begin{align*}
    \gamma &= \beta \theta + (1-\beta) \tau.
\end{align*}
As a result, we have
\begin{align*}
    &\beta \eta(B) + (1-\beta) \eta(C)
    \\ \geq& \beta (\theta \eta(A) + (1-\theta)\eta(D)) +
    \\&\quad\quad\quad\quad\quad\quad (1 -\beta) (\tau \eta(A) + (1-\tau) \eta(D))
    \\=& \gamma \eta(A) + (1-\gamma)\eta(D),
\end{align*}
indicating that Lemma \ref{lemma: concave on a line} is correct.
\end{IEEEproof}

Lemma \ref{lemma: concave on a line} is indeed a well-known result for concave function.
We now use it to simplify the proof of Theorem \ref{theorem: line to reduce complexity}.
For $1 \leq r \leq s \leq N$, denote
\begin{align*}
    a(r, s) &= \sum_{j = r}^{s} P_Y(y_j),
    \\b(r, s) &= \sum_{j = r}^{s} \frac{P_Y(y_j)}{a(r, s)} \delta_j.
\end{align*}
Then, according to \eqref{eqn: w}, we have
\begin{equation}\label{eqn: w = a b}
    w(r, s) = a(r, s) \phi(b(r, s)).
\end{equation}
Suppose $\delta_1, \delta_2, \ldots, \delta_N$ are sequentially located on a line.
In such a case, for any $1 \leq r < r' \leq s < s' \leq N$, $b(r, s), b(r, s'), b(r', s), b(r', s')$ are located on the line with $b(r, s)$ and $b(r', s')$ being the endpoints, and $\phi$ is concave on the line.
Let $\gamma = a(r, s) / (a(r, s) + a(r', s'))$ and $\beta = a(r, s') / (a(r, s') + a(r', s))$.
We have $\gamma, \beta \in [0, 1]$ and $\gamma b(r, s) + (1-\gamma)b(r', s') = \beta b(r, s') + (1-\beta) b(r', s)$.
By applying \eqref{eqn: w = a b} and Lemma \ref{lemma: concave on a line}, we have
\begin{align*}
    & (w(r, s) + w(r', s')) / (a(r, s) + a(r', s'))
    \\=& \gamma \phi(b(r, s)) + (1 - \gamma) \phi(r', s')
    \\\leq& \beta \phi(b(r, s')) + (1 - \beta) \phi(b(r', s))
    \\=& (w(r, s') + w(r', s)) / (a(r, s') + a(r', s)),
\end{align*}
leading to $w(r, s) + w(r', s') \leq w(r, s') + w(r', s) $.
This completes the proof.

\section{Proof of Theorem \ref{theorem: QI vs LLR increasing}}\label{appendix: proof of QI vs LLR increasing}

Our goal is to prove
\begin{equation}\label{eqn: w-alpha in appendix}
    w_{\alpha}(r, s) + w_{\alpha}(r', s') - w_{\alpha}(r, s') - w_{\alpha}(r', s) \leq 0
\end{equation}
for $\alpha \in (0, \infty]$ and for all $1 \leq r < r' \leq s < s' \leq N$ given that the elements in $\mc{X}$ can be relabelled to make $P_{Y|X}$ satisfy \eqref{eqn: (i, j) > (i', j')}.
Since $w_{\alpha}(\cdot, \cdot)$ is independent from the labelling of the elements of $\mc{X}$, for convenience, we assume that the elements in $\mc{X}$ has  been relabelled to make $P_{Y|X}$ satisfy \eqref{eqn: (i, j) > (i', j')}.

For any $\mb{a} = (a_i)_{1 \leq i \leq q}, \mb{b} = (b_i)_{1 \leq i \leq q} \in \mathbb{R}_+^q$, we use the following notations:
\begin{enumerate}[i)]
    \item   $\|\mb{a}\|_1 = \sum_{i = 1}^{q} a_i$,
    \item   $I_{min}(\mb{a}) = \arg \min_i (a_i = 0)$,
    \item   $I_{max}(\mb{a}) = \arg \max_i (a_i = 0)$,
    \item   $\mb{a} + \mb{b} = (a_i + b_i)_{1 \leq i \leq q}$,
\end{enumerate}
The proof is divided into four parts based on the four cases of $\alpha = 1$, $\alpha \in (0, 1)$,  $\alpha \in (1, \infty)$, and $\alpha = \infty$.

\emph{Part I: $\alpha = 1$}

Denote $\mb{p}, \mb{a}, \mb{b}, \mb{c} \in \mathbb{R}_+^q$, with $p_i, a_i, b_i, c_i$ given by
\begin{align}\label{eqn: p, a, b, c}
    p_i &= P_X(x_i),\nonumber\\
    a_i &= \sum_{j = r}^{r'-1} p_i P_{Y|X}(y_j|x_i),\nonumber\\
    b_i &= \sum_{j = r'}^{s} p_i P_{Y|X}(y_j|x_i),\nonumber\\
    c_i &= \sum_{j = s+1}^{s'} p_i P_{Y|X}(y_j|x_i).
\end{align}
Given \eqref{eqn: (i, j) > (i', j')}, we have
\begin{equation}\label{eqn: a > b > c}
    \mb{a} \succeq \mb{b}, \mb{b} \succeq \mb{c}, \mb{a} \succeq \mb{c},
\end{equation}
where $\succeq$ is defined in \eqref{eqn: >=}.
From \eqref{eqn: (i, j) > (i', j')} and \eqref{eqn: a > b > c}, we can easily derive
\begin{equation}\label{eqn: Imin Imax}
\left\{
    \begin{array}{l}
        I_{min}(\mb{a}) \leq I_{min}(\mb{b}) \leq I_{max}(\mb{a}) \leq I_{max}(\mb{b}), \\
        I_{min}(\mb{b}) \leq I_{min}(\mb{c}) \leq I_{max}(\mb{b}) \leq I_{max}(\mb{c}), \\
        u_i > 0, \forall \mb{u} \in \{\mb{a}, \mb{b}, \mb{c}\}, i \in [I_{min}(\mb{u}), I_{max}(\mb{u})].
    \end{array}
\right.
\end{equation}

For any $\mb{u} \in \mathbb{R}_+^q$, define
\[
    g(\mb{u}) = \sum_{i=1}^{q} u_i \log \frac{\| \mb{u} \|_1}{u_i},
\]
where we let $u_i \log \frac{\| \mb{u} \|_1}{u_i} = 0$ if $u_i = 0$.
Here the natural logarithm in base $e$ is used.
For other bases, the following proof can be similarly carried out.
In addition, let
\begin{align*}
    f(\mb{a}, \mb{b}, \mb{c}) &= g(\mb{a} + \mb{b}) + g(\mb{b} + \mb{c}) - g(\mb{a} + \mb{b} + \mb{c}) - g(\mb{b})
    \\&= w_{\alpha}(r, s) + w_{\alpha}(r', s') - w_{\alpha}(r, s') - w_{\alpha}(r', s).
\end{align*}
To prove $f(\mb{a}, \mb{b}, \mb{c}) \leq 0$, our idea is to properly modify $\mb{a}, \mb{b}$, and $\mb{c}$ in a series of steps, where after each step,  $f(\mb{a}, \mb{b}, \mb{c})$ keeps nondecreasing and finally becomes zero.
We summarize the procedure in Algorithm \ref{algo: modify a, b, c}, following which we also provide the remarks.

\begin{algorithm}[h!]
\caption{A series of modifications on $\mb{a}, \mb{b}$, and $\mb{c}$ such that $f(\mb{a}, \mb{b}, \mb{c})$ keeps nondecreasing and finally becomes zero}
\label{algo: modify a, b, c}
\begin{algorithmic}[1]
\REQUIRE $\mb{a}, \mb{b}, \mb{c}$ given by \eqref{eqn: p, a, b, c}.

\STATE  $k \leftarrow 1$.
\WHILE {$k < q$ and $\mb{a} \neq \mb{0}$}
    \IF {$a_k b_{k+1} > a_{k+1} b_k$}
        \STATE  $a_i \leftarrow a_{k+1} b_i / b_{k+1}, \forall i \in [k]$.\label{code: f @ ai}
    \ENDIF

    \IF {$b_k c_{k+1} > b_{k+1} c_k$}
        \IF {$b_{k+1} > 0$}
            \STATE  $c_i \leftarrow b_i c_{k+1}  / b_{k+1}, \forall i \in [k]$.\label{code: f @ ci > 0}
        \ELSE
            \STATE  $c_i \leftarrow 0, \forall i = k+1, k+2, \ldots, q$.\label{code: f @ ci = 0}
        \ENDIF
    \ENDIF
    \STATE  $k \leftarrow k + 1$.\label{code: f @ k+1}
\ENDWHILE
\STATE  $//$ At this point, we have $f(\mb{a}, \mb{b}, \mb{c}) = 0$.\label{code: f @ f = 0}
\end{algorithmic}
\end{algorithm}

\begin{remark}\label{remark: k = 1}
Note that for $k = 1$, \eqref{eqn: a > b > c}, \eqref{eqn: Imin Imax}, and the following conditions hold:
\begin{equation}\label{eqn: a = b = c}
    a_i b_j = a_j b_i, b_i c_j = b_j c_i, a_i c_j = a_j c_i, \forall 1 \leq i < j \leq k.
\end{equation}
Inductively, suppose these conditions (\eqref{eqn: a > b > c}--\eqref{eqn: a = b = c}) hold for $k < q$.
In the subsequent remarks, we will prove that these conditions can keep $f(\mb{a}, \mb{b}, \mb{c})$ nondecreasing after any modification of those in lines \ref{code: f @ ai}, \ref{code: f @ ci > 0}, \ref{code: f @ ci = 0} made to $\mb{a}, \mb{b}, \mb{c}$.
We will also prove that when Algorithm \ref{algo: modify a, b, c} reaches line \ref{code: f @ k+1} and increases $k$ by 1, either $\mb{a} = 0$ or these conditions will still hold.
It can be easily verified that either $\mb{a} = 0$ or these conditions that hold for $k = q$ can lead to $f(\mb{a}, \mb{b}, \mb{c}) = 0$ at line \ref{code: f @ f = 0}.
\end{remark}

\begin{remark}[for line \ref{code: f @ ai}]
Throughout this remark, let $k, \mb{a}, \mb{b}, \mb{c}$ refer to those at the beginning of line \ref{code: f @ ai} (before the modification).
Let $\mb{a}'$ refer to the $\mb{a}$ at the end of line \ref{code: f @ ai} (after the modification).
Our goal is to prove
\begin{equation}\label{eqn: f(a', b, c)}
    f(\mb{a}, \mb{b}, \mb{c}) \leq f(\mb{a}', \mb{b}, \mb{c}),
\end{equation}
i.e., to prove $f(\mb{a}, \mb{b}, \mb{c})$ keeps nondecreasing after the modification in line \ref{code: f @ ai}.

Let $T = a_{k+1} b_k / (a_k b_{k+1}) $.
For any $i \in [k]$, according to \eqref{eqn: a = b = c}, we have $T a_i b_k = T a_k b_i = a'_i b_k$.
This leads to $T a_i = a'_i$;
otherwise, we can easily derive a contradiction for $T a_i \neq a'_i$.
Let $t \in [T, 1]$ be a variable.
Denote $\mb{a}{(t)} = \left(a_i{(t)}\right)_{1 \leq i \leq q}$ with
\[
    a_i{(t)} =
    \begin{cases}
        a_i t, & i \in [k],\\
        a_i, & i \notin [k].
    \end{cases}
\]
Then, we have   $\mb{a}(1) = \mb{a}$ and $\mb{a}(T) = \mb{a}'$.

We are now to prove
\begin{equation}\label{eqn: at > b}
    \mb{a}(t) \succeq \mb{b}, \mb{a}(t) \succeq \mb{c}, \forall t \in [T, 1].
\end{equation}
For any $1 \leq i < j \leq k$ or $k < i < j \leq q$,
we can  easily verify $a_i(t) b_j \geq a_j(t) b_i$ according to \eqref{eqn: a > b > c}.
For $1 \leq i \leq k < j \leq q$, if $a_j(t) b_i = 0$, we also have $a_i(t) b_j \geq a_j(t) b_i$.
If $a_j(t) b_i > 0$, according to \eqref{eqn: a > b > c} and \eqref{eqn: Imin Imax}, we have $b_l > 0, \forall l \in [i, j]$, leading to $a_i(t) / b_i \geq a_{k+1} / b_{k+1} \geq a_j / b_j$.
Thus, $\mb{a}(t) \succeq \mb{b}$ holds.
To prove $\mb{a}(t) \succeq \mb{c}$, similarly, we only need to prove $a_i(t) c_j \geq a_j(t) c_i$ for $1 \leq i \leq k < j \leq q$ and $a_j(t) c_i > 0$, for which  we also have $b_j, c_j > 0$ according to \eqref{eqn: a > b > c} and \eqref{eqn: Imin Imax}.
Additionally, since $\mb{a}(t) \succeq \mb{b}$ holds, we have $a_i(t) / a_j(t) \geq b_i / b_{j} \geq c_i / c_j$.
This completes the proof of \eqref{eqn: at > b}.
Moreover, according to \eqref{eqn: a = b = c}, we have
\begin{align}\label{eqn: at = c}
    a_i(t) b_j = a_j(t) b_i, ~&a_i(t) c_j = a_j(t) c_i, \nonumber
    \\& \forall t \in [T, 1], 1 \leq i < j \leq k.
\end{align}

For $t \in (T, 1)$, we have
\begin{align}\label{eqn: partial f(a', b, c)}
    &\frac{\partial f(\mb{a}(t), \mb{b}, \mb{c})}{\partial t}\nonumber
    \\=& \sum_{i \in [k], a_i > 0} a_i \left( \log \frac{\|\mb{a}(t) + \mb{b}\|_1}{a_i(t) + b_i} -  \log \frac{\|\mb{a}(t) + \mb{b} + \mb{c}\|_1}{a_i(t) + b_i + c_i} \right)\nonumber\\
    \leq &0,
\end{align}
where the last inequality is due to
$
    \frac{\|\mb{a}(t) + \mb{b}\|_1}{a_i(t) + b_i} \leq \frac{\|\mb{a}(t) + \mb{b} + \mb{c}\|_1}{a_i(t) + b_i + c_i}
$
based on \eqref{eqn: a > b > c}, \eqref{eqn: a = b = c}, \eqref{eqn: at > b}, and \eqref{eqn: at = c}.
Moreover, since $f(\mb{a}(t), \mb{b}, \mb{c})$ is continuous at $t \in [T, 1]$, we have $f(\mb{a}, \mb{b}, \mb{c}) = f(\mb{a}(1), \mb{b}, \mb{c}) \leq f(\mb{a}(T), \mb{b}, \mb{c}) = f(\mb{a}', \mb{b}, \mb{c})$, leading to \eqref{eqn: f(a', b, c)}.

If $k = I_{max}(\mb{a}) < I_{max}(\mb{b})$, we have $\mb{a}' = \mb{0}$, in which case $f(\mb{a}', \mb{b}, \mb{c}) = 0$ holds and  the proof of part I is completed.
For $k < I_{max}(\mb{a})$ or $I_{max}(\mb{a}) = I_{max}(\mb{b})$, we always have $\mb{a'} \neq \mb{0}$.
After replacing $\mb{a}$ by $\mb{a}'$, \eqref{eqn: a > b > c}--\eqref{eqn: a = b = c} still hold according to \eqref{eqn: at > b} and \eqref{eqn: at = c}, except that $I_{min}(\mb{a})$ and $I_{max}(\mb{a})$ in \eqref{eqn: Imin Imax} are undefined for the case of $\mb{a} = \mb{0}$.
However, this exception does not affect the correctness of the proof of Part I.
\end{remark}

\begin{remark}[for line \ref{code: f @ ci > 0}]
Throughout this remark, let $k, \mb{a}, \mb{b}, \mb{c}$ refer to those at the beginning of line \ref{code: f @ ci > 0}.
Let $\mb{c}'$ refer to the $\mb{c}$ at the end of line \ref{code: f @ ci > 0}.
Our goal is to prove
\begin{equation}\label{eqn: f(a, b, c')}
    f(\mb{a}, \mb{b}, \mb{c}) \leq f(\mb{a}, \mb{b}, \mb{c}').
\end{equation}

Let $T = b_{k+1} c_k  / ( b_k c_{k+1}) $.
For any $i \in [k]$, according to \eqref{eqn: a = b = c}, we have $c_i b_k = c_k b_i = T c'_i  b_k$.
This leads to $c_i = T c'_i$ due to $b_k > 0$.
Let $t \in [T, 1]$ be a variable.
Denote $\mb{c}'{(t)} = \left(c'_i{(t)}\right)_{1 \leq i \leq q}$ with
\[
    c'_i{(t)} =
    \begin{cases}
        c'_i t, & i \in [k],\\
        c'_i, & i \notin [k].
    \end{cases}
\]
Then, we have   $\mb{c}'(1) = \mb{c}'$ and $\mb{c}'(T) = \mb{c}$.

Similar to \eqref{eqn: at > b}, we have
\begin{equation}\label{eqn: b > c't}
    \mb{b} \succeq \mb{c}'(t), \mb{a} \succeq \mb{c}'(t), \forall t \in [T, 1].
\end{equation}
Meanwhile, similar to \eqref{eqn: at = c}, we have
\begin{align}\label{eqn: b = c't}
    a_i c'_j(t) = a_j c'_i(t), ~&b_i c'_j(t) = b_j c'_i(t), \nonumber
    \\& \forall t \in [T, 1], 1 \leq i < j \leq k.
\end{align}
Then, for $t \in (T, 1)$, we have
\begin{align}\label{eqn: partial f(a, b, c')}
    &\frac{\partial f(\mb{a}, \mb{b}, \mb{c}'(t))}{\partial t}\nonumber
    \\=& \sum_{i \in [k], c'_i > 0} c'_i \left( \log \frac{\|\mb{b} + \mb{c}'(t)\|_1}{b_i + c'_i(t)} -  \log \frac{\|\mb{a} + \mb{b} + \mb{c}'(t)\|_1}{a_i + b_i + c'_i(t)} \right)\nonumber\\
    \geq &0,
\end{align}
where the last inequality is due to
$
    \frac{\|\mb{b} + \mb{c}'(t)\|_1}{b_i + c'_i(t)} \geq \frac{\|\mb{a} + \mb{b} + \mb{c}'(t)\|_1}{a_i + b_i + c'_i(t)}
$
based on \eqref{eqn: a > b > c}, \eqref{eqn: a = b = c}, \eqref{eqn: b > c't}, and \eqref{eqn: b = c't}.
Moreover, since $f(\mb{a}, \mb{b}, \mb{c}'(t))$ is continuous at $t \in [T, 1]$, we have $f(\mb{a}, \mb{b}, \mb{c}) = f(\mb{a}, \mb{b}, \mb{c}'(T)) \leq f(\mb{a}, \mb{b}, \mb{c}'(1)) = f(\mb{a}, \mb{b}, \mb{c}')$, leading to \eqref{eqn: f(a, b, c')}.
In addition, after replacing $\mb{c}$ by $\mb{c}'$, \eqref{eqn: a > b > c}--\eqref{eqn: a = b = c} still hold according to \eqref{eqn: b > c't} and \eqref{eqn: b = c't}.

\end{remark}

\begin{remark}[for line \ref{code: f @ ci = 0}]
Throughout this remark, let $k, \mb{a}, \mb{b}, \mb{c}$ refer to those at the beginning of line \ref{code: f @ ci = 0}.
Let $\mb{c}^*$ refer to the $\mb{c}$ at the end of line \ref{code: f @ ci = 0}.
Our goal is to prove
\begin{equation}\label{eqn: f(a, b, c^*)}
    f(\mb{a}, \mb{b}, \mb{c}) \leq f(\mb{a}, \mb{b}, \mb{c}^*),
\end{equation}

Let $t \in [0, 1]$ be a variable.
Denote $\mb{c}{(t)} = \left(c_i{(t)}\right)_{1 \leq i \leq q}$ with
\[
    c_i{(t)} =
    \begin{cases}
        c_i, & i \in [k],\\
        c_i t, & i \notin [k].
    \end{cases}
\]
Then, we have   $\mb{c}(0) = \mb{c}^*$ and $\mb{c}(1) = \mb{c}$.
Note that if Algorithm \ref{algo: modify a, b, c} reaches line \ref{code: f @ ci = 0}, we must have $k = I_{max}(\mb{b})$ according to \eqref{eqn: Imin Imax}.
As a result, we have $a_i = b_i = 0, \forall i = k+1, k+2, \ldots, q$.
Then, for $t \in (0, 1)$, we have
\begin{align}\label{eqn: partial f(a, b, c^*)}
    &\frac{\partial f(\mb{a}, \mb{b}, \mb{c}(t))}{\partial t}\nonumber\\
    =& \sum_{k < i \leq q, c_i > 0} c_i \left( \log \frac{\|\mb{b} + \mb{c}(t)\|_1}{c_i(t)} -  \log \frac{\|\mb{a} + \mb{b} + \mb{c}(t)\|_1}{c_i(t)} \right)\nonumber\\
    \leq &0.
\end{align}
Since $f(\mb{a}, \mb{b}, \mb{c}(t))$ is continuous at $t \in [0, 1]$, we have $f(\mb{a}, \mb{b}, \mb{c}) = f(\mb{a}, \mb{b}, \mb{c}(1)) \leq f(\mb{a}, \mb{b}, \mb{c}(0)) = f(\mb{a}, \mb{b}, \mb{c}^*)$, leading to \eqref{eqn: f(a, b, c^*)}.
In addition, after replacing $\mb{c}$ by $\mb{c}^*$, it can be easily verified that \eqref{eqn: a > b > c}--\eqref{eqn: a = b = c} still hold.

\end{remark}

\begin{remark}[for line \ref{code: f @ k+1}]
Let $\mb{a}, \mb{b}, \mb{c}$ refer to those at the beginning of line \ref{code: f @ k+1}.
Let $v$ be the value of $k$  at the end of line \ref{code: f @ k+1}.
If $\mb{a} = \mb{0}$, the proof of this part is indeed completed.
Suppose $\mb{a} \neq \mb{0}$.
Our final task is to prove that \eqref{eqn: a > b > c}--\eqref{eqn: a = b = c} still hold for $k = v$, since these conditions will be used for $k = v$ for the proof of \eqref{eqn: f(a', b, c)}, \eqref{eqn: f(a, b, c')}, and \eqref{eqn: f(a, b, c^*)}.

In the previous remarks, we have proved that \eqref{eqn: a > b > c}--\eqref{eqn: a = b = c} hold for $k = v-1$, no matter the modifications in lines \ref{code: f @ ai}, \ref{code: f @ ci > 0}, and \ref{code: f @ ci = 0} have been made or not.
As a result, \eqref{eqn: a > b > c} and \eqref{eqn: Imin Imax} still hold for $k = v$.
In order to prove \eqref{eqn: a = b = c}  for $k = v$, our task becomes to  prove
\begin{align}\label{eqn: a = b = c, k+1}
    a_i b_{v} = a_{v} b_i, b_i c_{v} = b_{v} c_i, a_i c_{v} &= a_{v} c_i, \forall i \in [v-1].
\end{align}
Note that when Algorithm \ref{algo: modify a, b, c} reaches line \ref{code: f @ k+1}, we always have $a_{v-1} b_{v} = a_{v} b_{v-1} \text{~and~} b_{v-1} c_{v} = b_{v} c_{v-1}$.
Based on this condition and that \eqref{eqn: a = b = c} holds for $k = v-1$, we can easily derive \eqref{eqn: a = b = c, k+1}.
At this point, the proof of Part I is completed.

\end{remark}

\emph{Part II: $\alpha \in (0, 1)$}

Denote $\mb{p}, \mb{a}, \mb{b}, \mb{c} \in \mathbb{R}_+^q$ by \eqref{eqn: p, a, b, c}.
For any $\mb{u} \in \mathbb{R}_+^q$, define
\begin{equation}\label{eqn: g for (0,1)}
    g(\mb{u}) = \left(\sum_{i=1}^{q} p_i^{1-\alpha} u_i^{\alpha} \right)^{1/\alpha}.
\end{equation}
In this case, we also have
\begin{align*}
    f(\mb{a}, \mb{b}, \mb{c}) &= g(\mb{a} + \mb{b}) + g(\mb{b} + \mb{c}) - g(\mb{a} + \mb{b} + \mb{c}) - g(\mb{b})
    \\&= w_{\alpha}(r, s) + w_{\alpha}(r', s') - w_{\alpha}(r, s') - w_{\alpha}(r', s).
\end{align*}
To prove $f(\mb{a}, \mb{b}, \mb{c}) \leq 0$, our idea is the same as that in Part I.
In this case, we indeed only need to prove \eqref{eqn: f(a', b, c)}, \eqref{eqn: f(a, b, c')}, and \eqref{eqn: f(a, b, c^*)} under the new definition of $g: \mathbb{R}_+^q \to \mathbb{R}$ given by \eqref{eqn: g for (0,1)}.
To this end, our task becomes to prove $\frac{\partial f}{\partial t} \leq 0, \frac{\partial f}{\partial t} \geq 0$, and $\frac{\partial f}{\partial t} \leq 0$ as what we do in \eqref{eqn: partial f(a', b, c)}, \eqref{eqn: partial f(a, b, c')}, and \eqref{eqn: partial f(a, b, c^*)}, respectively.
We complete these proofs below, where the notations correspond to those in \eqref{eqn: partial f(a', b, c)}, \eqref{eqn: partial f(a, b, c')}, and \eqref{eqn: partial f(a, b, c^*)}, except that $g$ is replaced by that defined by \eqref{eqn: g for (0,1)}.

\emph{Proof of $\frac{\partial f}{\partial t} \leq 0$ corresponding to \eqref{eqn: partial f(a', b, c)}:} We have
\begin{align*}
    &\frac{\partial f(\mb{a}(t), \mb{b}, \mb{c})}{\partial t}
    \\=& \sum_{i \in [k], a_i > 0} p_i^{1-\alpha} a_i  \left( \left( \sum_{j=1}^{q} p_j^{1-\alpha} \left(\frac{a_j(t) + b_j}{a_i(t) + b_i}\right)^\alpha \right)^{\frac{1-\alpha}{ \alpha}} - \right.\\
    &\quad\quad\quad \left. \left( \sum_{j=1}^{q} p_j^{1-\alpha} \left(\frac{a_j(t) + b_j + c_j}{a_i(t) + b_i + c_i}\right)^\alpha \right)^{\frac{1-\alpha}{ \alpha}} \right)\\
    \leq &0,
\end{align*}
where the last inequality is due to
$
    \frac{a_j(t) + b_j}{a_i(t) + b_i} \leq \frac{a_j(t) + b_j + c_j}{a_i(t) + b_i + c_i}
$
based on \eqref{eqn: a > b > c}, \eqref{eqn: a = b = c}, \eqref{eqn: at > b}, and \eqref{eqn: at = c}.

\emph{Proof of $\frac{\partial f}{\partial t} \leq 0$ corresponding to \eqref{eqn: partial f(a, b, c')}:} We have
\begin{align*}
    &\frac{\partial f(\mb{a}, \mb{b}, \mb{c}'(t))}{\partial t}\\
    =& \sum_{i \in [k], c'_i > 0} p_i^{1-\alpha} c'_i  \left( \left( \sum_{j=1}^{q} p_j^{1-\alpha} \left(\frac{b_j + c'_j(t)}{b_i + c'_i(t)}\right)^\alpha \right)^{\frac{1-\alpha}{ \alpha}} - \right.\\
    &\quad\quad\quad \left. \left( \sum_{j=1}^{q} p_j^{1-\alpha} \left(\frac{a_j + b_j + c'_j(t)}{a_i + b_i + c'_i(t)}\right)^\alpha \right)^{\frac{1-\alpha}{ \alpha}} \right)\\
    \geq &0,
\end{align*}
where the last inequality is due to
$
    \frac{b_j + c'_j(t)}{b_i + c'_i(t)} \geq \frac{a_j + b_j + c'_j(t)}{a_i + b_i + c'_i(t)}
$
based on \eqref{eqn: a > b > c}, \eqref{eqn: a = b = c}, \eqref{eqn: b > c't}, and \eqref{eqn: b = c't}.

\emph{Proof of $\frac{\partial f}{\partial t} \leq 0$ corresponding to \eqref{eqn: partial f(a, b, c^*)}:} We have
\begin{align*}
    &\frac{\partial f(\mb{a}, \mb{b}, \mb{c}(t))}{\partial t}\\
    =& \sum_{k < i \leq q, c_i > 0} p_i^{1-\alpha} c_i  \left( \left( \sum_{j=1}^{q} p_j^{1-\alpha} \left(\frac{b_j + c_j(t)}{ c_i(t)}\right)^\alpha \right)^{\frac{1-\alpha}{ \alpha}} - \right.\\
    &\quad\quad\quad \left. \left( \sum_{j=1}^{q} p_j^{1-\alpha} \left(\frac{a_j + b_j + c_j(t)}{c_i(t)}\right)^\alpha \right)^{\frac{1-\alpha}{ \alpha}} \right)\\
    \leq &0.
\end{align*}

\emph{Part III: $\alpha \in (1, \infty)$}

We omit the proof in this part since it can be carried out almost the same as that in Part II for $\alpha \in (0, 1)$.

\emph{Part IV: $\alpha = \infty$}

Set $\mb{a}, \mb{b}, \mb{c}$ with $a_i, b_i, c_i$ given by
\begin{align*}
    a_i &= \sum_{j = r}^{r'-1} P_{Y|X}(y_j|x_i),\\
    b_i &= \sum_{j = r'}^{s}  P_{Y|X}(y_j|x_i),\\
    c_i &= \sum_{j = s+1}^{s'} P_{Y|X}(y_j|x_i).
\end{align*}
In this case, \eqref{eqn: a > b > c} still holds.
Let $i = \arg \max_{1 \leq t \leq q}(a_t + b_t), j = \arg \max_{1 \leq t \leq q}(b_t + c_t), k = \arg \max_{1 \leq t \leq q}(a_t + b_t + c_t)$, and $l = \arg \max_{1 \leq t \leq q}b_t$.
Then, we have
\begin{align*}
    &w_{\alpha}(r, s) + w_{\alpha}(r', s') - w_{\alpha}(r, s') - w_{\alpha}(r', s)\\
    =& - (a_i + b_i) - (b_j + c_j) + (a_k + b_k + c_k) + b_l\\
    =& - (a_i + b_i) - (b_j + c_j) + (a_k + b_k) + \\
    &\,\,\,\quad\quad\quad\quad\quad\quad\quad\quad\quad\quad(b_l + c_l) + (c_k - c_l)\\
    \leq& c_k - c_l.
\end{align*}
Based on a similar deduction, we indeed have $w_{\alpha}(r, s) + w_{\alpha}(r', s') - w_{\alpha}(r, s') - w_{\alpha}(r', s) \leq \min\{ a_k - a_l, b_l - b_k, c_k - c_l \}$.
If $\min\{ a_k - a_l, b_l - b_k, c_k - c_l \} > 0$, we have both $a_k b_l > a_l b_k$ and $b_k c_l < b_l c_k$, leading to a contradiction to \eqref{eqn: a > b > c}.
Therefore, we must have $\min\{ a_k - a_l, b_l - b_k, c_k - c_l \} \leq 0$, implying \eqref{eqn: w-alpha in appendix} is true.

%


%
%

\ifCLASSOPTIONcaptionsoff
  \newpage
\fi

\bibliographystyle{IEEEtran}
\bibliography{myreference}

\begin{thebibliography}{10}
\providecommand{\url}[1]{#1}
\csname url@samestyle\endcsname
\providecommand{\newblock}{\relax}
\providecommand{\bibinfo}[2]{#2}
\providecommand{\BIBentrySTDinterwordspacing}{\spaceskip=0pt\relax}
\providecommand{\BIBentryALTinterwordstretchfactor}{4}
\providecommand{\BIBentryALTinterwordspacing}{\spaceskip=\fontdimen2\font plus
\BIBentryALTinterwordstretchfactor\fontdimen3\font minus
  \fontdimen4\font\relax}
\providecommand{\BIBforeignlanguage}[2]{{%
\expandafter\ifx\csname l@#1\endcsname\relax
\typeout{** WARNING: IEEEtran.bst: No hyphenation pattern has been}%
\typeout{** loaded for the language `#1'. Using the pattern for}%
\typeout{** the default language instead.}%
\else
\language=\csname l@#1\endcsname
\fi
#2}}
\providecommand{\BIBdecl}{\relax}
\BIBdecl

\bibitem{he2019dynamicISIT}
X.~He, K.~Cai, W.~Song, and Z.~Mei, ``Dynamic programming for quantization of
  $q$-ary input discrete memoryless channels,'' in \emph{Proc. IEEE Int. Symp.
  Inf. Theory}, Jul. 2019, pp. 450--454.

\bibitem{introAlgo01}
T.~H. Cormen, C.~E. Leiserson, R.~L. Rivest, and C.~Stein, \emph{Introduction
  to Algorithms: 2nd Edition}.\hskip 1em plus 0.5em minus 0.4em\relax
  Cambridge, MA, USA: MIT Press, 2001.

\bibitem{Kurkoski14}
B.~M. Kurkoski and H.~Yagi, ``Quantization of binary-input discrete memoryless
  channels,'' \emph{IEEE Trans. Inf. Theory}, vol.~60, no.~8, pp. 4544--4552,
  Aug. 2014.

\bibitem{Iwata14}
K.~Iwata and S.~Ozawa, ``Quantizer design for outputs of binary-input discrete
  memoryless channels using {SMAWK} algorithm,'' in \emph{Proc. IEEE Int. Symp.
  Inf. Theory}, Jun. 2014, pp. 191--195.

\bibitem{Sakai17}
Y.~Sakai and K.~Iwata, ``Optimal quantization of {B-DMCs} maximizing
  $\alpha$-mutual information with monge property,'' in \emph{Proc. IEEE Int.
  Symp. Inf. Theory}, Jun. 2017, pp. 2668--2672.

\bibitem{Aggarwal87}
A.~Aggarwal, M.~M. Klawe, S.~Moran, P.~Shor, and R.~Wilber, ``Geometric
  applications of a matrix-searching algorithm,'' \emph{Algorithmica}, vol.~2,
  no.~1, pp. 195--208, Nov. 1987.

\bibitem{Nazer17}
B.~Nazer, O.~Ordentlich, and Y.~Polyanskiy, ``Information-distilling
  quantizers,'' in \emph{Proc. IEEE Int. Symp. Inf. Theory}, Jun. 2017, pp.
  96--100.

\bibitem{Laber18}
E.~Laber, M.~Molinaro, and F.~M. Pereira, ``Binary partitions with approximate
  minimum impurity,'' in \emph{Proc. 35th Int. Conf. Machine Learning},
  vol.~80, Stockholmsm\"{a}ssan, Stockholm Sweden, Jul. 2018, pp. 2854--2862.

\bibitem{Burshtein92}
D.~Burshtein, V.~Della~Pietra, D.~Kanevsky, and A.~N\'{a}das, ``Minimum
  impurity partitions,'' \emph{Ann. Statist.}, vol.~20, no.~3, pp. 1637--1646,
  Sep. 1992.

\bibitem{coppersmith1999partitioning}
D.~Coppersmith, S.~J. Hong, and J.~R. Hosking, ``Partitioning nominal
  attributes in decision trees,'' \emph{Data Mining and Knowledge Discovery},
  vol.~3, no.~2, pp. 197--217, Jun. 1999.

\bibitem{Kurkoski08}
B.~M. Kurkoski, K.~Yamaguchi, and K.~Kobayashi, ``Noise thresholds for discrete
  {LDPC} decoding mappings,'' in \emph{Proc. IEEE Global Commun. Conf.}, Dec.
  2008, pp. 1--5.

\bibitem{Sakai14}
Y.~Sakai and K.~Iwata, ``Suboptimal quantizer design for outputs of discrete
  memoryless channels with a finite-input alphabet,'' in \emph{Proc. IEEE Int.
  Symp. Inf. Theory and Its Applications}, Nov. 2014, pp. 120--124.

\bibitem{Zhang16}
J.~A. Zhang and B.~M. Kurkoski, ``Low-complexity quantization of discrete
  memoryless channels,'' in \emph{Proc. IEEE Int. Symp. Inf. Theory and Its
  Applications}, Oct. 2016, pp. 448--452.

\bibitem{Hassanpour17}
S.~Hassanpour, D.~Wuebben, and A.~Dekorsy, ``Overview and investigation of
  algorithms for the information bottleneck method,'' in \emph{Proc. 11th Int.
  ITG Conf. on Systems, Commun. and Coding}, Feb. 2017, pp. 1--6.

\bibitem{lewandowsky2017message}
J.~Lewandowsky, M.~Stark, and G.~Bauch, ``Message alignment for discrete {LDPC}
  decoders with quadrature amplitude modulation,'' in \emph{Proc. IEEE Int.
  Symp. Inf. Theory}, Jun. 2017, pp. 2925--2929.

\bibitem{aslam2016read}
C.~A. Aslam, Y.~L. Guan, and K.~Cai, ``Read and write voltage signal
  optimization for multi-level-cell {(MLC) NAND} flash memory,'' \emph{IEEE
  Trans. Commun.}, vol.~64, no.~4, pp. 1613--1623, Feb. 2016.

\bibitem{mei2019onchannel}
Z.~Mei, K.~Cai, L.~Shi, and X.~He, ``On channel quantization for spin-torque
  transfer magnetic random access memory,'' \emph{IEEE Trans. Commun.},
  vol.~67, no.~11, pp. 7526--7539, Nov. 2019.

\bibitem{mei2020deep}
Z.~Mei, K.~Cai, and X.~He, ``Deep learning-aided dynamic read thresholds design
  for multi-level-cell flash memories,'' \emph{IEEE Trans. Commun.}, vol.~68,
  no.~5, pp. 2850--2862, May 2020.

\bibitem{Wang14}
J.~Wang, K.~Vakilinia, T.-Y. Chen, T.~Courtade, G.~Dong, T.~Zhang, H.~Shankar,
  and R.~Wesel, ``Enhanced precision through multiple reads for {LDPC} decoding
  in flash memories,'' \emph{IEEE J. Sel. Areas Commun.}, vol.~32, no.~5, pp.
  880--891, May 2014.

\bibitem{he2019mutual}
\BIBentryALTinterwordspacing
X.~He, K.~Cai, and Z.~Mei, ``Mutual information-maximizing quantized belief
  propagation decoding of regular {LDPC} codes,'' \emph{arXiv}, 2019. [Online].
  Available: \url{https://arxiv.org/abs/1904.06666}
\BIBentrySTDinterwordspacing

\bibitem{he2019onfinite}
X.~{He}, K.~{Cai}, and Z.~{Mei}, ``On finite alphabet iterative decoding of
  {LDPC} codes with high-order modulation,'' \emph{IEEE Commun. Lett.},
  vol.~23, no.~11, pp. 1913--1917, Nov. 2019.

\bibitem{Yao80}
F.~F. Yao, ``Efficient dynamic programming using quadrangle inequalities,'' in
  \emph{Proc. 12th Annual ACM Symposium on Theory of Computing}.\hskip 1em plus
  0.5em minus 0.4em\relax New York, NY, USA: ACM, 1980, pp. 429--435.

\bibitem{Verdu15}
S.~Verd{\'u}, ``$\alpha$-mutual information,'' in \emph{Proc. IEEE Information
  Theory and Applications Workshop}, Feb. 2015, pp. 1--6.

\bibitem{Ho15}
S.~Ho and S.~Verd{\'u}, ``Convexity/concavity of {R\'{e}nyi} entropy and
  $\alpha$-mutual information,'' in \emph{Proc. IEEE Int. Symp. Inf. Theory},
  Jun. 2015, pp. 745--749.

\bibitem{Csiszar95}
I.~Csisz{\'a}r, ``Generalized cutoff rates and {R\'{e}nyi}'s information
  measures,'' \emph{IEEE Trans. Inf. Theory}, vol.~41, no.~1, pp. 26--34, Jan.
  1995.

\bibitem{Bein09}
W.~Bein, M.~J. Golin, L.~L. Larmore, and Y.~Zhang, ``The {Knuth-Yao}
  quadrangle-inequality speedup is a consequence of total monotonicity,''
  \emph{ACM Trans. Algorithms}, vol.~6, no.~1, pp. 17:1--17:22, Dec. 2009.

\bibitem{gallager1965simple}
R.~Gallager, ``A simple derivation of the coding theorem and some
  applications,'' \emph{IEEE Trans. Inf. Theory}, vol.~11, no.~1, pp. 3--18,
  Jan. 1965.

\bibitem{polyanskiy2010arimoto}
Y.~Polyanskiy and S.~Verd{\'u}, ``Arimoto channel coding converse and
  {R{\'e}nyi} divergence,'' in \emph{Proc. Annual Allerton Conference on
  Communication, Control, and Computing}, Sep. 2010, pp. 1327--1333.

\end{thebibliography}

\end{document}